\documentclass[12pt]{article}
\usepackage{amsmath, amssymb, graphics}
\usepackage{floatflt}
\usepackage{algorithm}
\usepackage{algorithmic}
\usepackage{wrapfig}
\usepackage{xcolor}
\usepackage[dvips]{epsfig}
\usepackage{amsfonts}
\usepackage{moreverb}
\usepackage{fullpage}
\usepackage{graphicx}
\usepackage{amsmath}

\newtheorem{theorem}{Theorem}[section]
\newtheorem{lemma}[theorem]{Lemma}
\newtheorem{observation}[theorem]{Observation}
\newtheorem{corollary}[theorem]{Corollary}
\newtheorem{proposition}[theorem]{Proposition}
\newtheorem{claim}{Claim}[section]

\def\Cost{\mbox{\tt Cost}}

\def\inline#1:{\par\vskip 7pt\noindent{\bf #1:}\hskip 10pt}
\def\Proof{\par\noindent{\bf Proof:~}}
\def\blackslug{\hbox{\hskip 1pt \vrule width 4pt height 8pt
    depth 1.5pt \hskip 1pt}}
\def\QED{\quad\blackslug\lower 8.5pt\null\par}

\newcommand{\R}{\mathbb{R}}

\newcommand{\MAXDEG}[0]{\Delta}
\newcommand{\DIFF}[0]{\mbox{Diff}}

\newcommand{\PS}[0]{$\mbox{\tt Secluded}$ $\mbox{\tt Steiner}$
$\mbox{\tt Tree}$}
\newcommand{\Terminals}[0]{\mathcal{S}}
\newcommand{\PP}[0]{$\mbox{\tt Secluded}$ $\mbox{\tt Path}$}
\newcommand{\DegThreeConst}[0]{4}
\newcommand{\DegAnyConst}[0]{\MAXDEG+1}
\newcommand{\MLP}[0]{\mbox{\tt Minimum Labeled Path}}
\newcommand{\MBP}[0]{\mbox{\tt Maximal Breach Path}}
\newcommand{\MLST}[0]{\mbox{\tt Minimum Labeled Spanning Tree}}
\newcommand{\MLC}[0]{\mbox{\tt Minimum Labeled Cut}}
\newcommand{\LPM}[0]{\mbox{\tt Labelled Prefect Matching}}
\newcommand{\SuperIndex}[0]{\widetilde{I}}
\newcommand{\TD}[0]{\mathcal{T}}
\newcommand{\Steiner}[0]{\mbox{\tt Steiner Tree}}
\newcommand{\NodeSteiner}[0]{\mbox{\tt Node} \mbox{\tt Weighted}
\mbox{\tt Steiner} \mbox{\tt Tree}}
\newcommand{\PSDeg}[0]{\mbox{\tt Min Degree Steiner Tree}}
\newcommand{\RBSC}[0]{$\mbox{\tt Red-Blue}$ $\mbox{\tt Set}$ $\mbox{\tt Cover}$}
\newcommand{\SC}[0]{\mbox{\tt Set Cover}}
\newcommand{\VC}[0]{\mbox{\tt Vertex Cover}}

\newcommand{\Suff}[0]{\mathrm{Suff}}

\newcommand{\Last}[0]{\mathrm{Last}}
\newcommand{\dist}{\mbox{\rm dist}}

\def\cG{{\cal G}}
\def\Cost{\mbox{\tt Cost}}
\def\DegCost{\mbox{\tt DegCost}}
\def\HighDeg{\MAXDEG^{*}}
\def\OptDegCost{\DegCost^{*}}

\title{Secluded Connectivity Problems}

\author{
Shiri Chechik
\thanks{
\hbox{Microsoft Research Silicon Valley Center, USA. Email:}
{\tt schechik@microsoft.com}}
\and
M. P. Johnson
\thanks{
\hbox{Department of Electrical Engineering, UCLA, Los Angeles, USA. Email:}
{\tt mpjohnson@gmail.com}}
\and
Merav Parter
\thanks{The Weizmann Institute of Science, Rehovot, Israel.
Email: {\tt \{merav.parter,david.peleg\}@ weizmann.ac.il}.}
\thanks{Recipient of the Google Europe Fellowship in distributed computing;
 research supported in part by this Google Fellowship.}
\and
David Peleg $^\ddagger$
\thanks{Supported in part by the Israel Science Foundation
(grant 894/09), the United States-Israel Binational Science Foundation
(grant 2008348), the Israel Ministry of Science and Technology
(infrastructures grant), and the Citi Foundation.}
} 
\begin{document}

\maketitle

\begin{abstract}
Consider a setting where possibly sensitive information sent over a path in a network is visible to every {neighbor} of the path, i.e., every neighbor of some node on the path, thus including the nodes {\em on} the path itself. The {\em exposure} of a path $P$ can be measured as the number of nodes adjacent to it, denoted by $N[P]$. A path is said to be {\em secluded} if its exposure is small. A similar measure can be applied to other connected subgraphs, such as Steiner trees connecting a given set of terminals. Such subgraphs may be relevant due to considerations of privacy, security or revenue maximization.
This paper considers problems related to minimum exposure connectivity structures such as paths and Steiner trees. It is shown that on unweighted undirected $n$-node graphs, the problem of finding the minimum exposure path connecting a given pair of vertices is strongly inapproximable, i.e., hard to approximate within a factor of $O(2^{\log^{1-\epsilon}n})$ for any $\epsilon>0$ (under an appropriate complexity assumption), but is approximable with ratio $\sqrt{\MAXDEG}+3$, where $\MAXDEG$ is the maximum degree in the graph. One of our main results concerns the class of bounded-degree graphs, which is shown to exhibit the following interesting dichotomy. On the one hand, the minimum exposure path problem is NP-hard on \emph{node-weighted} or \emph{directed} bounded-degree graphs (even when the maximum degree is 4). On the other hand, we present a polynomial algorithm (based on a nontrivial dynamic program) for the problem on unweighted undirected bounded-degree graphs. Likewise, the problem is shown to be polynomial also for the class of  (weighted or unweighted) bounded-treewidth graphs.
Turning to the more general problem of finding a minimum exposure Steiner tree connecting a given set of $k$ terminals, the picture becomes more involved. In undirected unweighted graphs with unbounded degree, we present an approximation algorithm with ratio $\min\{\MAXDEG, n/k, \sqrt{2n},O(\log k \cdot (k+\sqrt{\MAXDEG}))\}$. On unweighted undirected bounded-degree graphs, the problem is still polynomial when the number of terminals is fixed, but if the number of terminals is arbitrary, then the problem becomes NP-hard again.
\end{abstract}

\section{Introduction}
\vskip .1cm \noindent \textbf{The problem.}
Consider a setting where possibly sensitive information sent over a path in a network is visible to every {neighbor} of the path, i.e., every neighbor of some node on the path, thus including the nodes {\em on} the path itself. The {\em exposure} of a path $P$ can be measured as the size (possibly node-weighted) of its \emph{neighborhood} in this sense, denoted by $N[P]$. A path is said to be {\em secluded} if its exposure is small. A similar measure can be applied to other connected subgraphs, such as Steiner trees connecting a given set of terminals.
Our interest is in finding connectivity structures with exposure as low as possible. This may be motivated by the fact that in real-life applications, a connectivity structure operates normally as part of the entire network  $G$ (and is not ``extracted'' from it), and so controlling the effect of its operation on the other nodes in the network may be of interest, in situations in which any ``activation'' of a node (by taking it as part of the structure) leads to an activation of its neighbors as well. In such settings, to minimize the set of total active nodes, we aim toward finding secluded or sufficiently private connectivity structures. Such subgraphs may be important in contexts where privacy is an important concern, or in settings where security measures must be installed on any node from which the information is visible, making it desirable to minimize their number. Another context  where minimizing exposure may be desirable is when the information transferred among the participants has commercial value and overexposure to ``free viewers'' implies revenue loss.

This paper considers the problem of minimizing the exposure of subgraphs that satisfy some desired connectivity requirements.  Two fundamental connectivity problems are considered, namely, single-path connectivity and Steiner trees, formulated as the \PP\ and \PS\ problems, respectively, as follows. Given a graph $G=(V,E)$ and an $s,t$ pair (respectively, a terminal set $\Terminals$), it is required to find an $s-t$ path (respectively, a Steiner tree) of minimum exposure.

\vskip .1cm \noindent \textbf{Related Work.}
The problems considered in this paper are variations of the classical
shortest path and Steiner tree problems. In the standard versions of
these problems, a cost measure is associated with edges or vertices, e.g., representing length or weight and the task is to identify a minimum cost subgraph satisfying the relevant connectivity requirement. Essentially, the cost of the solution subgraph is a \emph{linear} sum of the solution's \emph{constituent parts}, i.e., the sum of the weights of the edges or vertices chosen.

In contrast, in the setting of \emph{labeled connectivity} problems, edges (and occasionally vertices) are associated with \emph{labels} (or \emph{colors}) and the objective is to identify a subgraph $G' \subseteq G$ that satisfies the connectivity requirements while minimizing the number of used labels. In other words, costs are now assigned to labels rather than to single edges. Such labeling schemes incorporate grouping constraints, based on partitioning the set of available edges into classes, each of which can be purchased in its entirety or not at all. These grouping constraints are motivated by applications from telecommunication networks, electrical networks, and multi-modal transportation networks. Labeled connectivity problems have been studies extensively from complexity-theoretic and algorithmic points of view \cite{DS99,YuanVJ05,HassinMS07,FellowsGK10}. The optimization problems in this category include, among others, the $\MLP$ problem \cite{HassinMS07,YuanVJ05}, the $\MLST$ problem \cite{KrumkeW98,HassinMS07}, the $\MLC$ problem \cite{ZhangCTZ11}, and the $\LPM$ problem \cite{Monnot05}.

In both the traditional setting and the labeled connectivity setting, only edges or nodes that are explicitly part of the selected output structure are ``paid for'' in solution cost. That is, the cost of a candidate structure is a pure function of its components, ignoring the possible effects of ``passive'' participants, such as nodes that are ``very close'' to the structure in the input graph $G$.  In contrast, in the setting considered in this paper, the cost of a connectivity structure $G'$ is a function not only of its components but also of their immediate surroundings, namely, the manner in which $G'$ is embedded in $G$ plays a role as well. (Alternatively, we can say that the cost is a not necessarily a linear function of its components.)

To the best of our knowledge, secluded connectivity problems have not been considered before in the literature. The \PP\ and \PS\  problems are related to several existing combinatorial optimization problems. These include the \RBSC\ problem \cite{CarrDKM00,Peleg07}, the $\MLP$  problem \cite{HassinMS07,YuanVJ05} and the $\Steiner$ \cite{Kar72} and \NodeSteiner\ problems \cite{KleinR95}. A prototypical example is the \RBSC\ problem, in which we are given a set $R$ of red elements, a set $B$ of blue elements and a family $S \subseteq 2^{|R|\cup |B|}$ of subsets of blue and red elements, and the objective is to find a subfamily $C \subseteq S$ covering all blue elements that minimizes the number of red elements covered. This problem is known to be strongly inapproximable.

Finally, turning to geometric settings, similarly motivated problems have been studied in the networking and sensor networks communities, where sensors are often modeled as unit disks. For example, the \MBP\ problem
\cite{MeguerdichianKPS01} is defined in the context of traversing a region
of the plane that contains sensor nodes at predetermined points,
and its objective is to maximize the minimum distance between the points
on the path and the the sensor nodes.
(The solution uses edges of the sensor nodes' Voronoi diagram.)
A dual problem studied extensively is {\em barrier coverage},
i.e., the (deterministic or stochastic) placement of sensors
in order to make it difficult for an adversary to cross the
region unseen (see \cite{LiuDWS08} and the references therein).
\cite{ChenKL10} studies $k$-barrier coverage, in which the task is to
position the sensors so that any path across the region will intersect
with at least $k$ sensor disks.
Similarly motivated problems have been studied in the context of path planning
in AI. ``Stealth'' path planning problems, in which the task is to find a
minimum ``visibility'' path from source to destination, have been considered in
\cite{Johansson:2010:KPM:1948395.1948440,MarzouqiJ06,MarzouqiJ11}.
Although the motivation is similar, such problems are technically
quite different from the graph-based problems studied here;
those problems are typically posed in the geometric plane, amid obstacles
that cause occlusion, and visibility is defined in terms of line-of-sight.
\vskip .1cm \noindent \textbf{Contributions.}
In this paper, we introduce the concept of \emph{secluded connectivity} and study some of its complexity and algorithmic aspects. We first show that the \PP\ (and hence also \PS) problem is strongly inapproximable on unweighted undirected graphs with unbounded degree (more specifically, is hard to approximate with ratio $O(2^{\log^{1-\epsilon}n})$, where $n$ is the number of nodes in the graph $G$, assuming $\mathcal{NP} \not\subseteq  \mathcal{DTIME}\left(n^{\text{poly}\log n} \right)$).
Conversely, we devise a $\sqrt{\MAXDEG}+3$ approximation algorithm for the \PP\ problem and a $\min\{\MAXDEG, n/k, \sqrt{2n}, O(\log k \cdot (k+\sqrt{\MAXDEG}))\}$ approximation algorithm for the \PS\ problem, where $\MAXDEG$ is the maximum degree in the graph and $k$ is the number of terminals.

One of our key results concerns bounded-degree graphs and reveals an interesting dichotomy. On the one hand, we show that \PP\ is NP-hard on the class of \emph{node-weighted} or \emph{directed} bounded-degree graphs, even if the maximum degree is 4. In contrast, we show that on the class of unweighted undirected bounded-degree graphs, the \PP\ problem admits an \emph{exact} polynomial-time algorithm, which is based on a complex dynamic programming and requires some nontrivial analysis. Likewise, the \PS\ problem with fixed size terminal set is in P as well.

Finally, we consider some specific graph classes. We show that the \PP\ and \PS\ problems are polynomial for
bounded-treewidth graphs. We also show that the \PP\ (resp., \PS) problem can be approximated with ratio $O(1)$ (resp., $\Theta(\log k)$) in polynomial time for hereditary graph classes of bounded density. As an example,
the \PP\ problem has a 6 approximation on planar graphs.
(A more careful direct analysis of the planar case yields ratio 3.)

\vskip .1cm \noindent \textbf{Preliminaries.}
Consider a node-weighted graph $G(V,E,W)$,
for some weight function $W: V \to \R_{\geq 0}$,
with $n$ nodes and maximum degree $\MAXDEG$.
For a node $u \in V$, let $N(u)=\{v \in V \mid (u,v) \in E\}$ be the set
of $u$'s neighbors and let $N[u]=N(u) \cup \{u\}$ be $u$'s
{\em closed neighborhood}, i.e.,
including $u$ itself.
A {\em path} is a sequence $P=[u_1, \ldots,u_\ell]$, oriented from left
to right, also termed a $u_1-u_\ell$ path.
Let $P[i]=u_i$ for $i \in \{1, \ldots, \ell\}$.
Let $First(P)=u_1$ and $Last(P)=u_\ell$.
For a path $P$ and nodes $x,y$ on it, let $P[x,y]$ be the  subpath of $P$
from $x$ to $y$. For a connected subgraph $G' \subseteq G$ and for
$u_i,u_j \in V(G')$, let $\dist_{G'}(u_i,u_j)$ be the distance between $u_i$
and $u_j$ in $G'$. Let $N(G')=\bigcup_{u \in G'} N(u) \setminus G'$ be the nodes that are strictly neighbors of $G'$ nodes and
$N[G']=\bigcup_{u \in G'} N[u]$ be the set of nodes in the 1-neighborhood of $G'$.
Define the cost of $G'$ as
\begin{equation}
\label{eq:wcost}
\Cost(G')=\sum_{u \in N[G']} W(u)~.
\end{equation}
Note that if $G$ is unweighted, then the cost of a subgraph $G'$ is simply
the cardinality of the set of $G'$ nodes and their neighbors,
$\Cost(G')=|N[G']|$.

We sometimes consider the neighbors of node $u \in V(G)$ in different subgraphs. To avoid confusion, we denote $N_{G'}(u)$ the neighbors of $u$ restricted to graph $G'$.

For a subgraph $G' \subseteq G$, let $\DegCost(G')$ denote the sum of the degrees of the nodes of $G'$. If $G'$ is a path, then this key parameter  is closely related to our problem. It is not hard to see that for any given path $P$, $\Cost(P) \leq \DegCost(P)$. The problem of finding an $s-t$ path $P$ with minimum $\DegCost(P)$ is polynomial, making it a convenient starting-point for various heuristics for the problem.

In this paper we consider two main connectivity problems. In the \PP\ problem we are given an unweighted graph $G(V,E)$, a source node $s$ and target node $t$, and the objective is to find an $s-t$ path $P$ with minimum neighborhood size.
A generalization of this problem is the \PS\ problem, in which instead of two terminals $s$ and $t$ we are given a set of $k$ terminal nodes $\Terminals$ and it is required to find a tree $T$ in $G$ covering $\Terminals$, of minimum neighborhood size.
If the given graph $G$ is weighted, then the \emph{weighted} \PP\ and \PS\ problems require minimizing the neighborhood cost as given in Eq. (\ref{eq:wcost}).
We now define these tasks formally. For an $s-t$ pair, let ${\cal P}_{s,t}=\{P \mid P \text{~is a s-t path}\}$ be the set of all $s-t$ paths and let $q_{s,t}^{*}=\min \{\Cost(P) \mid P \in {\cal P}_{s,t}\}$ be the minimum cost among these paths. Then the objective of the \PP\ problem is to find a $P^{*}\in {\cal P}_{s,t}$ that attains this minimum, i.e., such that $\Cost(P^{*})=q^{*}_{s,t}$.
For the \PS\ problem, let ${\cal T}(\Terminals)=\{T \subseteq G \mid \Terminals \subseteq V(T), \text{$T$ is a tree}\}$ be the set of all trees in $G$ covering $\Terminals$, and let $q^{*}(\Terminals)$ be the minimum cost among these trees, i.e.,
\begin{equation}
\label{eq:opt_cost}
q^{*}(\Terminals)=\min \{\Cost(T) \mid T \in {\cal T}(\Terminals)\}~.
\end{equation}
Then the solution for the
problem is a tree $T^{*} \in {\cal T}(\Terminals)$ such that $\Cost(T^{*})= q^{*}(\Terminals)$.

%
\vskip .1cm \noindent \textbf{Notation}
For a graph $G=(V,E)$ and a set $\Terminals$ of $k$ terminals, let $T$ be some Steiner tree (spanning all terminals). Consider a decomposition of the tree into $O(k)$ subpaths $\Sigma(T)$ defined as follows.
Let $B^{*}(T)=\Terminals \cup \{u \in V(G) \mid |N_T(u)|\geq 3\}$ be the set of terminals and nodes with more than three neighbors in $T$ (thus branching points in $T$). For $a,b \in B^{*}(T)$, let $\alpha=[a,b]$ be some maximal path in $T$ such that $V(\alpha) \cap B^{*}(T)=\{a,b\}$, i.e., $\alpha$ is a maximal subpath in $T$ whose internal section is \emph{free} of $B^{*}(T)$ nodes. Since any internal node in $\alpha$ is of degree at most 2, $\alpha$ is indeed a path in $T$. Let $\Sigma(T)=\{\alpha=[a,b] \mid a,b \in B^{*}(T)\}$ be the collection of these maximal subpaths. Note that $|\Sigma(T)|\leq 2k$.

\section{Unweighted Undirected Graphs with Unbounded Degree}

\begin{figure}[h!]
\begin{center}
\includegraphics[scale=0.3]{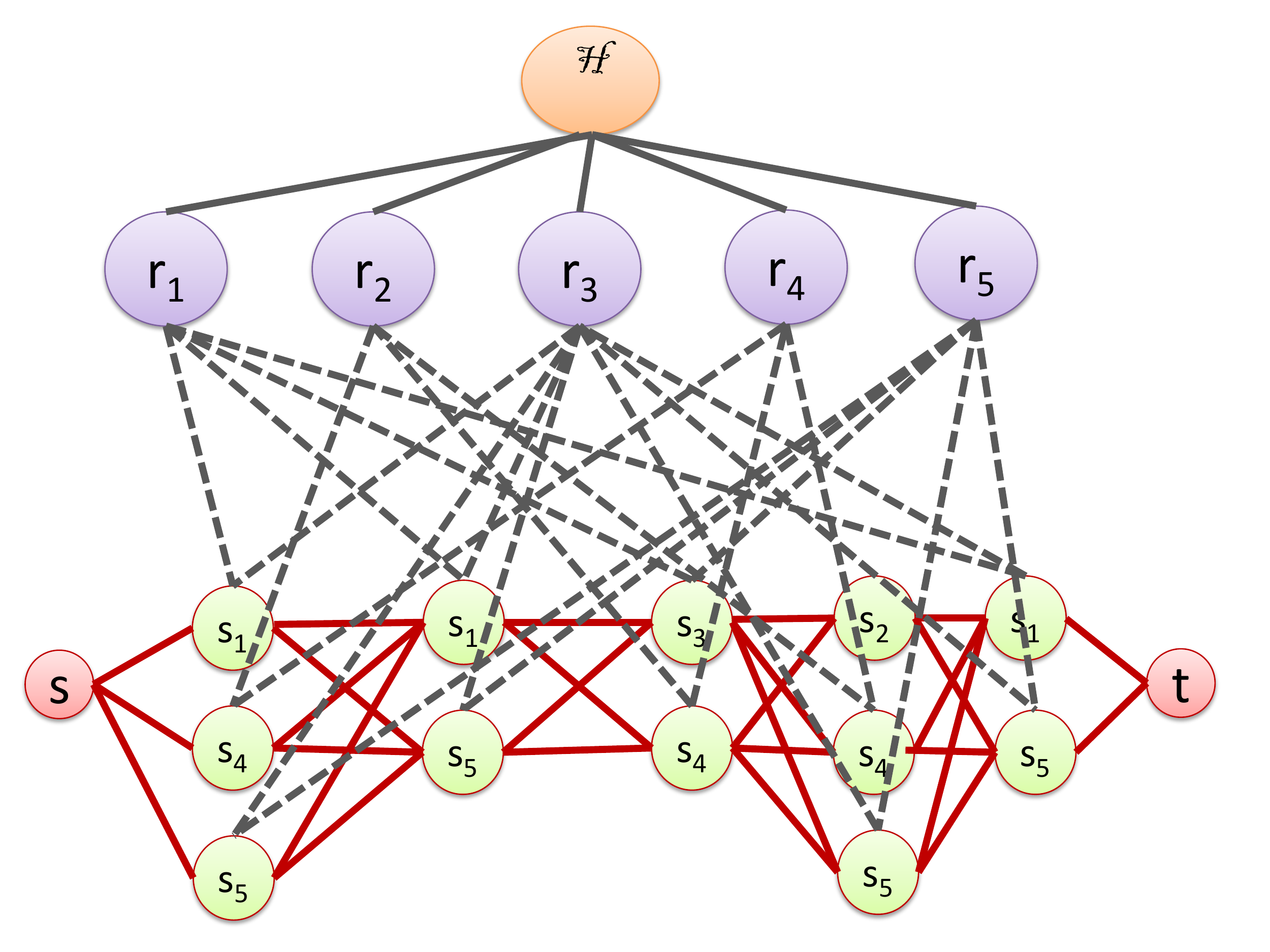}
\caption{ \label{fig:hardness}
\sf
Example of the gap-preserving reduction from \RBSC\ to \PP.
The \RBSC\ instance is given by $S_1=\{b_1,b_2,b_3,r_1,r_3\}, S_2=\{b_4, r_4\}, S_3=\{b_3, r_1,r_5\}, S_4=\{b_1,b_3,b_4,r_2,r_4\}, S_5=\{b_1,b_2,b_4,b_5,r_3,r_5 \}$.  The optimal solution for \RBSC\ is $\{S_1, S_3, S_5\}$, which covers $3$ red elements. The corresponding $s-t$ path is of cost $14+3 \cdot 5^{3}=389$. Note that an alternative solution of $\{S_1, S_4\}$ is of higher cost, covering 4 red elements, and the corresponding $s-t$ path is of cost $14+4 \cdot 5^{3}=514$.
}
\end{center}
\end{figure}

\vskip .1cm \noindent \textbf{Hardness of approximation.}
\begin{theorem}
\label{thm:hardness}
On unweighted undirected graphs with unbounded degree, the \PP\ problem (and hence also the \PS\ problem) is strongly inapproximable. Specifically, unless $\mathcal{NP} \subseteq  \mathcal{DTIME}(n^{poly\log(n)})$, the \PP\ problem cannot be approximated to within a factor $O(2^{\log^{1-\epsilon}n})$ for any $\epsilon>0$.
\end{theorem}
We show this for the simple case where $\Terminals=\{s,t\}$. The theorem is shown by a gap-preserving reduction from the \RBSC\ (RBSC) problem \cite{Peleg07}. An instance of \RBSC\ consists of a set $B=\{b_1, \ldots, b_{|B|}\}$ of blue elements, a set $R=\{r_1, \ldots, r_{|R|}\}$ of red elements, and a collection $S=\{S_1, \ldots, S_m\}$ of subsets of $B\cup R$. The task is to choose a family of sets covering all blue elements but a minimum number of red elements. In \cite{CarrDKM00} it is shown that the \RBSC\ problem cannot be approximated to within an $O(2^{\log^{1-\epsilon} n})$ ratio unless $\mathcal{NP} \subseteq  \mathcal{DTIME}(n^{poly\log(n)})$, where $n=\max\{|R|,|B|,|S|\}$.
Given a \RBSC\ instance, we construct a \PP\ instance $(G,s,t)$ as follows.
For each $b_i$, let $\mathcal{S}_{i}=\{S_j \in S \mid b_i \in S_j\}$ be the sets containing $b_i$.
For every $i, j$, add to $G$ a node $v_{i,j}$ corresponding to $S_j \in \mathcal{S}_{i}$.
Define $\mathcal{V}_{i}=\bigcup_{S_j \in \mathcal{S}_{i}} \{v_{i,j}\}$ and
$\mathcal{V}=\bigcup_{i} \mathcal{V}_i$.
Add edges so that the sets $\mathcal{V}_{i}$ and $\mathcal{V}_{i+1}$ form a complete bipartite graph for every $1 \le i < |B|$.
In addition, we connect a node $s$ to all nodes in $\mathcal{V}_{1}$, and connect a node $t$ to all nodes in $\mathcal{V}_{|B|}$.
So far, our construction contains $X=\sum_{i} |\mathcal{V}_{i}|+2 \leq n^{2}+2$ nodes (see Fig. \ref{fig:hardness}).

We now turn to describing the representation of the red elements $R$. For every $r_\ell \in R$, let $\mathcal{C}_{\ell}$ be a ``supernode'' consisting of $n^{3}$ nodes (with no edges among them), and let $\mathcal{C}=\bigcup_{\ell=1}^{|R|}\mathcal{C}_{\ell}$ be the set of all supernodes. All $n^{3}$ nodes of $\mathcal{C}_{\ell}$ are connected to node $v_{i,j}$ if $r_\ell \in S_j$, for every $\ell \in \{1, \ldots, |R|\}$, $i \in \{1, \ldots, |B|\}$, and $j  \in \{1, \ldots, |S|\}$.  Finally, let $\mathcal{H}$ be a ``hypernode'' consisting of $n^{5}$ nodes (with no edges among them), all of which are connected to all the constituent nodes of all supernodes.
This completes the construction of $G$. Overall, $|V(G)|=O(n^{5})$.

By ``visiting'' a supernode or the hypernode, we mean visiting any of their constituent nodes.
We now make an immediate observation which allows us to restrict ourselves to $s-t$ paths $P$ not visiting any supernode or the hypernode, i.e., with $V(P) \subseteq \mathcal{V}$.
\begin{claim}
\label{cl:approx_1_n}
No $\omega(1/n)$-approximate secluded path solution will visit any supernodes or the hypernode, i.e., will satisfy $V(P) \cap \mathcal{C}=\emptyset$.
\end{claim}
\Proof
For every path $P$ that goes through a node $v \in \mathcal{C} \cup \mathcal{H}$ it holds that $\Cost(P) \geq n^{5}$. On the other hand, a path $P'$ that visits only $\mathcal{V}$ nodes, i.e., $V(P') \subseteq \mathcal{V}$ costs only $\Cost(P') \leq X+|R|\cdot n^{3}=O(n^{4})$. The claim follows.
\QED

Let $\SuperIndex(P)= \{ i \mid \mathcal{C}_i \subseteq N[P]\}$ be the indices of the supernodes in the neighborhood of path $P$. Then
\begin{equation*}
\Cost(P)=X+|\SuperIndex(P)| \cdot n^{3}.
\end{equation*}
We first show the correctness of the reduction and then consider gap-preservation.
\begin{claim}
\label{cl:rbcs_pp}
There exists an $s-t$ path of cost $X+kn^3$ iff there exists an \RBSC\ solution of cost $k$.
\end{claim}
\Proof
($\Rightarrow$) By Claim \ref{cl:approx_1_n}, we restrict ourselves to paths not visiting any supernodes. Thus, by the structure of the graph, any such $s-t$ path visits one of the nodes $v_{i,j}$ for each $1 \le i \le |B|$, corresponding to the selection of $S_j$ to cover $b_i$. Since all $X$ ordinary nodes will be in $N(P)$, $kn^3$ of the cost is due to the size-$n^3$ supernodes, which correspond to $k$ red elements covered in the \RBSC\ solution.

($\Leftarrow$) Conversely, given a cost-$k$ \RBSC\ solution, we construct the path as follows. Starting from $s$, for every $1 \le i \le |B|$, add to $P$ a node $v_{i,j}$ such that $S_j$ is in the \RBSC\ solution, and then finally add $t$. The cost due to the ordinary nodes is $X$ and due to the $k$ supernodes is $kn^3$.
\QED

We now show that the reduction is gap-preserving.
By Claim \ref{cl:rbcs_pp} it follows that $q^{*}_{s,t}=X+|R^{*}| \cdot n^{3}$, where $|R^*|$ is the optimal \RBSC\ cost (the number of red elements $R^{*}$ covered by the optimal solution).
Assume that there exists an $\alpha>O(1/n)$ approximation algorithm for the \PS\ problem. This would result in path a $P$ such that
\begin{equation*}
\Cost(P)= X+|\SuperIndex(P)| \cdot n^{3}\leq \alpha(X+|R^{*}| \cdot n^{3})\leq 2\alpha \cdot |R^{*}| \cdot n^{3}~.
\end{equation*}
Thus $|\SuperIndex(P)| \leq 2\alpha |R^{*}|$, implying that
$(\Cost(P)-X)/n^{3}$ is a $2\alpha$ approximation to \RBSC. Since $|V(G)|=n^{5}$, and as \RBSC\ is inapproximable within a factor of $O(2^{\log^{1-\epsilon}n})$ for every fixed $\epsilon$, we get that \PP\ is inapproximable within a factor of $O(2^{\log^{1-\epsilon'}n})$, where $\epsilon'=f(\epsilon)$ for some appropriate function $f$.
This complete the proof of Thm. \ref{thm:hardness}
\QED
\begin{corollary}
\label{cor:dag}
The \PP\ problem (and hence also the \PS\ problem) is strongly inapproximable in directed acyclic graphs.
\end{corollary}
\Proof
The proof is by gap-preserving reduction from \RBSC, very similar to the general case. The  key modifications from the general case are that the hypernode of $n^{5}$ nodes is not included and we add directions to the edges; directing the edges from $s$ toward the nodes of $S_1$, namely, $\mathcal{V}_{1}$, from the nodes of $\mathcal{V}_{i}$ to the nodes of $\mathcal{V}_{i+1}$ for every $i \in \{1, \ldots, |B|\}$, and from the nodes of $\mathcal{V}_{|B|}$ to the node $t$. We also direct all the supernodes edges towards the supernode elements. It can be verified that this directed graph is acyclic. The rest of the analysis is as in the undirected case. The graph $G$ contains $O(n^{3})$ nodes, thus the \PP\ problem cannot be approximated within an $O(2^{\log^{1-\epsilon}n})$ ratio. The corollary follows.
\QED

\subsection{Approximation}
\begin{theorem}
\label{thm:delta_approx}
The \PP\ problem in unweighted undirected graphs can be approximated within a ratio of $\sqrt{\MAXDEG}+3$.
\end{theorem}
\Proof
Given an instance of the \PP\ problem, let $P^*$ be an $s-t$ path that minimizes $\Cost(P^{*})$. Note that we may assume without loss of generality that for every node $u$ in $P^*$, the only neighbors of $u$ in $G$ from among the nodes of $V(P^{*})$ are the nodes adjacent to $u$ in $P^*$. To see this, note that otherwise, if $u$ had an edges to some neighbor $u' \in V(P^{*})$ such that $e$ is not on $P^{*}$, we could have shortened the path $P^{*}$ (by replacing the subpath from $u$ to $u'$ with the edge $e$) and obtained a shorter path with at most the same cost as $P^*$.

Recall that $\DegCost(P)$ denotes the sum of the node degrees of the path $P$, and that $\Cost(P) \leq \DegCost(P)$ for any $P$. Recall also that the problem of finding an $s-t$ path $P$ with minimum $\DegCost(P)$ is polynomial. We claim that the algorithm that returns the path $Q^*$ minimizing $\DegCost(Q^*)$ yields a $(\sqrt{\MAXDEG}+3)$ approximation ratio for the \PP\ problem.

In order to prove this, we show there exists a path $Q$ such that $\DegCost(Q) \leq (\sqrt{\MAXDEG}+3) \cdot \Cost(P^{*})$.
This implies that $\DegCost(Q^*) \leq \DegCost(Q) \leq (\sqrt{\MAXDEG}+3) \cdot \Cost(P^{*})$, as required.

The path $Q$ is constructed by the following iterative process.
Initially, all nodes are unmarked and we set $Q =P^*$.
While there exists a node with more than $\sqrt{\MAXDEG}+3$ unmarked neighbors on the path $Q$, pick such a node $x$.
Let $y$ be the first (closest to $s$) neighbor of $x$ in $Q$ and let $z$ be the last (closest to $t$) neighbor of $x$ in $Q$.
Replace the subpath $Q[y,z]$ in $Q$ with the path $[y,x,z]$ and mark the node $x$.
We now show that $\DegCost(Q) \leq (\sqrt{\MAXDEG}+3) \cdot \Cost(P^{*})$.
Let $X=\{x_1, \ldots x_\ell\}$ be the set of marked nodes in $Q$, where $x_i$ is the node marked in iteration $i$.
Note that $\ell=|X|$ is the number of iterations in the entire process.

For the sake of analysis, partition $N[P^*]$ into two sets: $S_1$, those that have no unmarked neighbor in $Q$, and $S_2$, those that do.
We claim that $|S_1| \geq \ell \cdot \sqrt{\MAXDEG}$.

For each $x_i \in X$, let $P(x_i)$ be the path that was replaced by the process of constructing $Q$ in iteration $i$, and let $Q(x_i)$ be the path $Q$ in the beginning of that iteration.
Since $x_i$ has more than $\sqrt{\MAXDEG} +3$ unmarked neighbors in $Q(x_i)$, we get that $|P(x_i) \cap P^*| \geq \sqrt{\MAXDEG} +4$.
Let $P^{-}(x_i)$ be the path obtained by removing the first two nodes and last two nodes from $P(x_i)$.
Note that $|P^{-}(x_i) \cap P^*| \geq \sqrt{\MAXDEG}$.
In addition, the sets $P^{-}(x_i) \cap P^*$ for $i=\{1, \ldots, \ell\}$ are pairwise disjoint, and moreover, none of the nodes in $Q$ has a neighbor in $P^{-}(x_i) \cap P^*$.
We thus get that $P^{-}(x_i) \cap P^* \subseteq S_1$.
Note that $\Cost(P^*) = |S_1| + |S_2|$ and that $\ell \leq  |S_1| /\sqrt{\MAXDEG}$.
We get that $\DegCost(Q) \leq |S_2| \cdot (\sqrt{\MAXDEG}+3) +
\ell \cdot \MAXDEG \leq |S_2| \cdot (\sqrt{\MAXDEG}+3) + |S_1| \cdot \sqrt{\MAXDEG} \leq (|S_2|+|S_1|) \cdot (\sqrt{\MAXDEG}+3) = \Cost(P^*)\cdot (\sqrt{\MAXDEG}+3)$.
\QED

In addition, for the \PS\ problem with $k$ terminals, we establish the following.
\begin{theorem}
\label{thm:steiner_delta_approx}
The \PS\ problem in unweighted undirected graphs can be approximated within a ratio of
$\min\{\MAXDEG, n/k, \sqrt{2n}, O(\log k \cdot (k+\sqrt{\MAXDEG}))\}.$
\end{theorem}
Recall that $\DegCost(T)$ is the sum of degrees in the tree $T$ and
let $\OptDegCost(\Terminals)=\min\{\DegCost(T) \mid T \subseteq G \text{~spans~} \Terminals\}$. Clearly, $\Cost(T)\leq \DegCost(T)$. \\The $\PSDeg$ problem is defined as follows. Given a graph $G=(V,E)$ and a set of terminals $\Terminals$, the objective is to find a Steiner tree $T$ of minimal $\DegCost(T)$, i.e., such that $\DegCost(T)=\OptDegCost(\Terminals)$.
\begin{claim}
\label{cl:degcost_approx}
The $\PSDeg$ problem is $NP$-hard for arbitrary $k=|\Terminals|$ and can be approximated within a ratio of $\Theta(\log k)$. If $k$ is constant, then $\PSDeg$ is polynomial.
\end{claim}
\Proof
We first present an $O(\log k)$ approximation algorithm.
In the \NodeSteiner\ problem, nonnegative costs are assigned to nodes as well as to edges. The cost of subgraph $G' \subseteq G$ that contains the terminals $\Terminals$ is the sum of the costs of its nodes and edges. While constant factor approximations exists for the standard $\Steiner$ problem (of edge-weighted graphs), the \NodeSteiner\ problem cannot be approximated to within less than a logarithmic factor $O(\log k)$ assuming that $NP \nsubseteq DTIME(n^{O(\log\log n)})$, by a reduction from $\SC$ \cite{Feige98,KleinR95}.
Using the $O(\log k)$-approximation algorithm of  \cite{Feige98}, the $\PSDeg$ problem can be solved as follows. Set the weight of each vertex to be its degree, that is $W(v)=|N(v)|$, for every $v \in V(G)$, and set zero weights to the edges.
The optimal \NodeSteiner\ tree with these weights corresponds to the optimal $\PSDeg$ tree. Since \NodeSteiner\ is polynomial for fixed $k$, so is $\PSDeg$ \cite{KleinR95}.

We now show that this approximation is tight.
For brevity, let $\Pi$ denote the special case of the \NodeSteiner\ problem in which edges have zero weights and $\Pi'$ denote the $\PSDeg$ task.
The approximation-preserving reduction from $\SC$ presented in \cite{KleinR95} assigns zero weights to the edges, and hence $\Pi$ is inapproximable within logarithmic factor.  By reducing from $\Pi$ to $\Pi'$ we will show that $\Pi'$ is inapproximable within logarithmic factor as well. Given a $\Pi$ instance graph $G(V,E,W)$ and terminal set $\Terminals$, where $W(v_i)\geq 0$, we transform it into an instance $G', \Terminals$ of $\Pi'$. We assume that the weights are polynomial. To obtain $G'$, the weights $W$ are scaled (multiplied by a common large enough factor $X$) to weights $W'$ such that $W'(v_i)\geq |N_G(v_i)|$ for every node $v_i \in V$. Next, define the residual of vertex $v_i$ as $Res(v_i)=W'(v_i)-|N_G(v_i)|$. The scaling step guarantees that $Res(v_i)\geq 0$.
Next, add to each vertex $v_i$ a disjoint set $\Upsilon(v_i)=\{ v_i^j, j \in [1, Res(v_i)]\}$ consisting of $1$-weight $Res(v_i)$ neighbors. Formally, $G'=(V',E',W')$, where $V'=V \cup_{v_i \in V} \Upsilon(v_i)$ and $E'=E \cup \{ (v_i, v_i^{j}), v_i^{j}\in \Upsilon(v_i), v_i \in V\}$. Finally,  $W'(v_i)=|N_{G'}(v_i)|$ for every $v_i \in V$ and $W'(u)=1$ for every $u \in V' \setminus V$. It is easy to see that optimal solution of $\Pi'$ in $G'$ corresponds to an optimal solution for $\Pi$ in $G$. Since the weights $W$ are polynomial (hence so is $X$) it follows that a $\Omega(\log k)$ approximation for  $\Pi'$ in $G'$ would imply an $\Omega(\log k)$ approximation for  $\Pi$ in $G$, in contradiction to the logarithmic inapproximability of $\Pi'$ due to \cite{KleinR95}.
\QED
We now show the following.
\begin{lemma}
\label{cl:degcost}
$\OptDegCost(\Terminals)=\Theta(\sqrt{\MAXDEG}+k)\cdot q^{*}(\Terminals)$.
\end{lemma}
\Proof
Let $T^{*}$ be an optimal secluded Steiner tree, meaning $\Cost(T^{*})=q^{*}(\Terminals)$. Starting with $T^{*}$ we will perform a set of operations on it, ending with a spanning $\Terminals$ tree $\widehat{T}$ for which $\DegCost(\widehat{T}) \leq \Theta(\sqrt{\MAXDEG}+k) \Cost(T^{*})$. Since $\DegCost(\widehat{T})\geq \OptDegCost(\Terminals)$ this would establish that $\OptDegCost(\Terminals)<O(\sqrt{\MAXDEG}+k)\cdot q^{*}(\Terminals)$. We begin by describing the procedure for obtaining $\widehat{T}$ from $T^{*}$ and then analyze its cost, i.e.,  $\DegCost(\widehat{T})$. Note that as we do not know $T^{*}$; this procedure is only theoretical to aid the analysis and is not meant for implementation.

Let $\HighDeg=\sqrt{\MAXDEG}+4k$. Throughout, we assume that $\MAXDEG> \HighDeg$. Else, any $O(1)$-approximation algorithm for the \Steiner\ problem is a $O(\MAXDEG)$-approximation algorithm to the \PS\ problem.
Initially, let all nodes in $T^{*}$ be unmarked and set $\widehat{T}=T^{*}$.
A node $u$ is \emph{high-degree with respect to} $\widehat{T}$ if it has at least $\HighDeg$ unmarked neighbors in $\widehat{T}$. While there is a high-degree node $x$ with respect to $\widehat{T}$, do the following. For each subpath $\alpha=[a,b] \in \Sigma(T^{*})$ such that $|N(x) \cap V(\alpha)|\geq 2$, i.e., $x$ has at least two neighbors in $\alpha$, replace the following subpaths. Let $y_1$ (respectively, $y_2$) be the neighbor of $x$ closest to $a$ (respectively, $b$) on $T^{*}$. Replace the subpath $[y_1,y_2] \subseteq \alpha$ with the subpath $[y_1,x,y_2]$ and mark the node $x$.

We now show that $\DegCost(\widehat{T}) \leq O(\sqrt{\MAXDEG}+k)q^{*}(\Terminals)$. Let $M=\{u_1, \ldots, u_y\}$ be the set of marked nodes in $\widehat{T}$. This implies that the procedure consists of $|M|=y$ iterations. Let $T_0=T^{*}$ and $T_i$ be the spanning $\Terminals$ subgraph at the end of iteration $i$ for $i=1, ..., |M|$ (thus $T_{y}=\widehat{T}$). The node $u_i \in M$ is the high-degree node with respect to $T_{i-1}$ that was observed at iteration $i$. Let $\overline{M}=V(\widehat{T}) \setminus M$ be the unmarked nodes in the final subgraph $\widehat{T}$. Note that $\widehat{T}$ might not be a tree (due to cycle introduced by the marked nodes), but clearly, a subgraph of it is a  tree spanning all terminals. By definition, the $\DegCost$ of $\widehat{T}$ is given by
\begin{equation}
\label{eq:degcost_t}
\DegCost(\widehat{T}) =\DegCost(M)+\DegCost(\overline{M})~,
\end{equation}
where $\DegCost(U)$, for $U \subseteq V$, is given by $\DegCost(U)=\sum_{u \in U}|N_G(u)|$. Let $M_i=\{u_j \mid j\leq i\}$ be the set of marked nodes at the end of iteration $i$. Given a path $P=[x_1, x_2, \ldots, x_{q-1}, x_q]$, let $P^{-}=[x_2, \ldots, x_{q-1}]$ denote its internal subpath. For every marked node $u_i$ and a subpath $\alpha \in \Sigma(T^{*})$ for which $u_i$ has at least $3$ unmarked neighbors in $\alpha$ at iteration $i$, let $P_i(\alpha)$ be the subpath in $T_{i-1}$ that was shortcut in $T_i$ by introducing $u_i$; otherwise let $P_i(\alpha)=\emptyset$.
Let $N^{-}(u_i)=N_{T_{i-1}}(u_i) \setminus M_{i-1}$ be the unmarked neighbors of $u_i$ at iteration $i$ and define
$\Psi(u_i)=\bigcup_{\alpha \in \Sigma(T^{*})}N^{-}(u_i) \cap V(P_i^{-}(\alpha))$, the set of ``internal'' unmarked neighbors of $u_i$ in $T_{i-1}$ (the nodes that are removed from $T_{i-1}$ due to path replacements).

We now make two observations. (A) $|\Psi(u_i)|\geq \sqrt{\MAXDEG}$ for every node $u_i \in M$, and (B) $\Psi(u_1),...,\Psi(u_{y})$ are pairwise disjoint. We start with (A). Note that for every $\alpha \in \Sigma(T^{*})$ on which $u_i$ has unmarked neighbors, we ignore at most $2$ nodes in $\Psi(u_i)$ (corresponding to the endpoints of the replaced path, $P_i(\alpha)$). Since there are at most $2k$ subpaths in $\Sigma(T^{*})$, there are at most $4k$ unmarked neighbors of $u$ in $T_{i-1}$ that are ignored in $\Psi(u_i)$. As $u_i$ has $\HighDeg$ unmarked neighbors in $T_{i-1}$, we get that
$|\Psi(u_i)| \geq \HighDeg-4k\geq \sqrt{\MAXDEG}$, which establishes (A). Note that (B) holds since each set $\Psi(u_i) \subseteq V(T_{i-1})$ is part of the tree in iteration $i$ but is not part of the tree in iteration $i+1$, i.e. $\Psi(u_i) \notin V(T_i)$ (due to the subpath replacement step).
%
We now refer to Eq. (\ref{eq:degcost_t}) and bound $\DegCost(M)$ and $\DegCost(\overline{M})$. First, observe that $\DegCost(M)\leq \sqrt{\MAXDEG} \cdot q^{*}(\Terminals)$. This holds since
every node $u_i \in M$ corresponds to a disjoint set $\Psi(u_i) \subseteq V(T^{*})$ of at least $\sqrt{\MAXDEG}$ nodes of the optimal tree $T^{*}$. Thus, $|M| \leq V(T^{*})/\sqrt{\MAXDEG}$ and $|M| \leq q^{*}(\Terminals)/\sqrt{\MAXDEG}$. Since the degree of each marked node is bounded by $\MAXDEG$, it follows that
\begin{equation}
\label{eq:degcost_M}
\DegCost(M) \leq \sqrt{\MAXDEG} \cdot q^{*}(\Terminals)~.
\end{equation}
We now bound the degree of the unmarked nodes $\widehat{M}$ of $\widehat{T}$. Let $N^{*}=N_G[\widehat{M}]$.
Since $\widehat{M} \subseteq V(T^{*})$, it follows that $N^{*} \subseteq N[T^{*}]$ and hence $|N^{*}| \leq q^{*}(\Terminals)$.
By the stopping criteria, it follows that any node in $N^{*}$ has at most $\HighDeg$ neighbors in $\widehat{M}$.
Hence
\begin{equation}
\label{eq:degcost_Mhat}
\DegCost(\widehat{M})\leq \HighDeg \cdot q^{*}(\Terminals).
\end{equation}

\begin{figure}[h!]
\begin{center}
\includegraphics[scale=0.3]{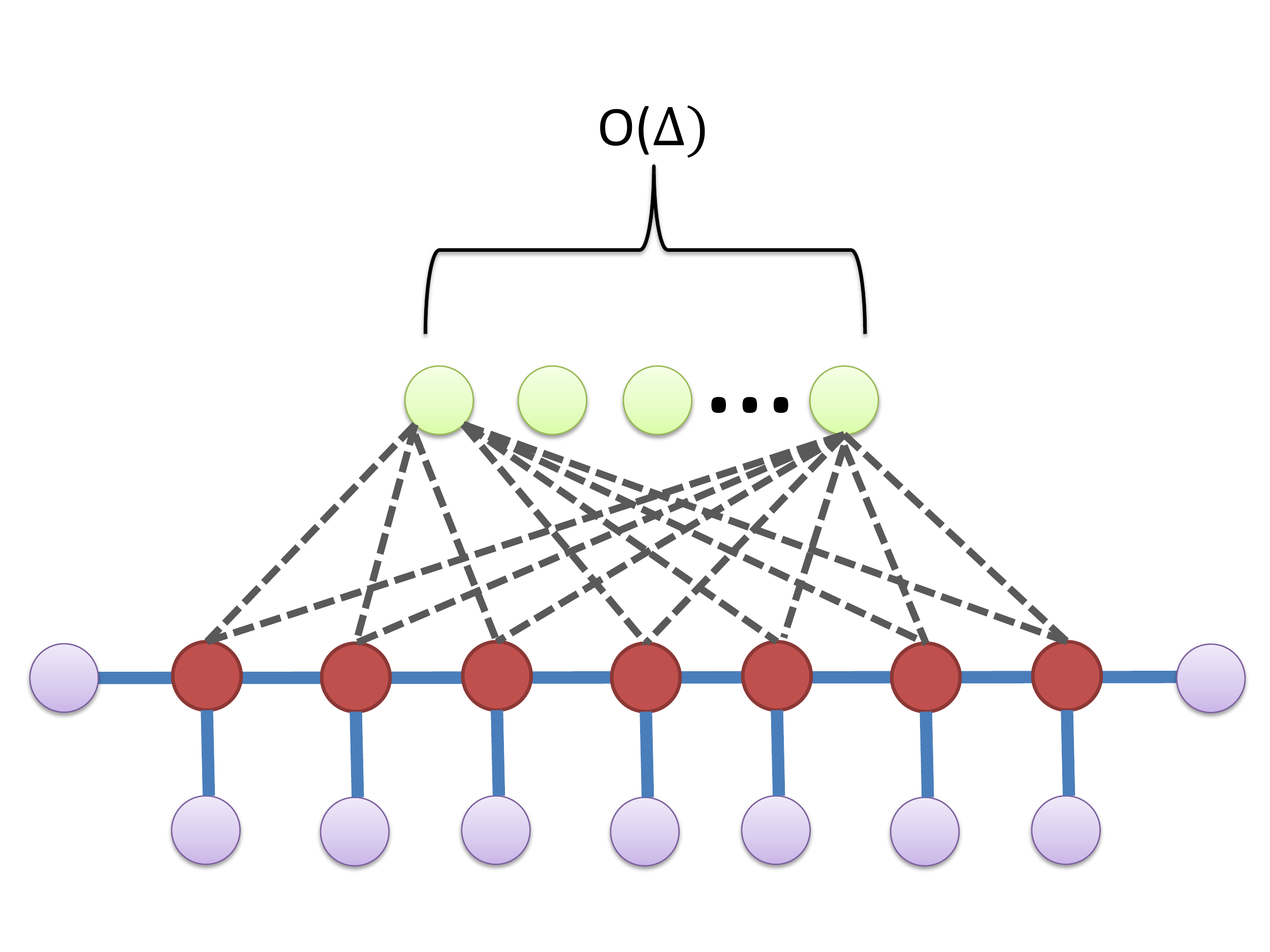}
\caption{ \label{fig:degcost_gap}
\sf
Illustration of the gap between $q^{*}(\Terminals)$ and $\DegCost^{*}(\Terminals)$.  The light purple nodes correspond to the set of terminals $\Terminals$.
}
\end{center}
\end{figure}

Combining Eqs. (\ref{eq:degcost_t}), (\ref{eq:degcost_M}), and (\ref{eq:degcost_Mhat}), it follows that $\OptDegCost(\Terminals)=O(\sqrt{\MAXDEG}+k)\cdot q^{*}(\Terminals)$.
Finally, we show that there exists a graph $G$ and a set of terminals $\Terminals$ for which $\OptDegCost(\Terminals)=\Omega(\sqrt{\MAXDEG}+k) \cdot q^{*}(\Terminals)$. Set $3<\MAXDEG<k$, graph $G$  consists of $2k+\MAXDEG-5$ nodes. Let $\Terminals=\{v_0, ..., v_{k-1}\}$. The skeleton for graph $G$ is a path $P=[v_0,u_1, \ldots, u_{k-2},v_{k-1}]$. Then, for every $i \in [1,k-2]$, connect $u_i$ to the terminal $v_i$. In addition, connect each of the $u_i$ nodes to the set of nodes $Z=\{z_1, \ldots, z_{\MAXDEG-3}\}$. The graph $G$ is presented in Fig. \ref{fig:degcost_gap}. Clearly, the optimal tree $T^{*}$ under both measures, i.e., $\Cost(T^{*})$ and $\DegCost(T^{*})$, is given by the subgraph induced by $G(V \setminus Z)$. Since every $Z$ node is counted $k-2$ times in $\DegCost(T^{*})$ instead of one time in $\Cost(T^{*})$, it follows that
$\DegCost(T^{*})=\MAXDEG \cdot (k-2)+k$ and $\Cost(T^{*})=2k+\MAXDEG-5$. Since $k\geq \MAXDEG$, we get that $\DegCost(T^{*})=\Omega(\sqrt{\MAXDEG}+k) \Cost(T^{*})$. The lemma follows.
\QED
\begin{claim}
\label{cl:steiner_sqrtn}
The \PS\ problem on unweighted undirected graphs can be approximated within $\sqrt{2n}$.
\end{claim}
\Proof
Let $g$ be a guess for $q^{*}(\Terminals)$, the solution of the optimal secluded Steiner tree. A Steiner tree $T_{g}$ is constructed for each for each guess $g \in \{1, \ldots, n\}$ as follows. Remove from $G$ all vertices with degree $\geq g$ and let $T_{g}$ be a $2$-approximation for the optimal Steiner tree (with minimum number of edges). Let $T'=T_{g^{*}}$ for $g^{*}=q^{*}(\Terminals)$, i.e., the 2-approximation Steiner tree (where the objective is to minimize the number of edges) computed when considering the correct guess. We show that $\Cost(T')\leq \sqrt{2n} \cdot q^{*}(\Terminals')$.  Let $L$ be the cost of the optimal Steiner tree (i.e., number of edges in the tree). Then clearly $q^{*}(\Terminals) \geq L$.
In addition, $\Cost(T') \leq 2L \cdot q^{*}(\Terminals)$. This follows, since our approximated Steiner tree $T'$ contains at most twice the edges of the optimal one, and the degree of each vertex is bounded by $q^{*}(\Terminals)$.
Finally, since the cost of any secluded Steiner tree is bounded by the number of nodes $n$, we get that $Cost(T')/q^{*}(\Terminals)=\min\{n/q^{*}(\Terminals), 2\cdot q^{*}(\Terminals)\} \leq \sqrt{2n}$.
\QED
We are now ready to complete the proof for Thm. \ref{thm:delta_approx}.
\Proof
First note that any $O(1)$-approximation algorithm to the \Steiner\ problem is an $O(\MAXDEG)$-approximation algorithm to \PS. Next, note that obtaining an $n/k$ approximation is trivial, since $\Cost(T) \leq n$ for any spanning tree $T$ and $q^{*}(\Terminals)\geq k$.
The approximation of $\sqrt{2n}$ is thanks to Claim \ref{cl:steiner_sqrtn} and that of $\Theta(\log k (\sqrt{\MAXDEG}+k))$ is due to Claims \ref{cl:degcost} and \ref{cl:degcost_approx}. Specifically, let $\mathcal{A}$ be an $\Theta(\log k)$ approximation algorithm for the $\PSDeg$ problem, and let $T$ be the resulting spanning $\Terminals$ tree output by algorithm $\mathcal{A}$ given graph $G$ and $\Terminals$ as input. Due to Claim \ref{cl:degcost_approx}, such an $\mathcal{A}$ exists, and hence $\DegCost(T)<\Theta(\log k) \OptDegCost(\Terminals)$. Combining this with \ref{cl:degcost}, we get that $\DegCost(T) \leq \Theta(\log k (\sqrt{\MAXDEG}+k))\cdot q^{*}(\Terminals)$, as required. The theorem follows.
\QED

\section{Bounded-Degree Graphs}
In this section we show that the ~\PP\ problem (and thus also the \PS\ problem) is $NP$-hard on both directed graphs and weighted graphs, even if the maximum node degree is $4$. We have the following.
\begin{lemma}
\label{lem:privatepath_vc}
The \PP\ problem is NP-complete even for graphs of maximum degree $4$
if they are either (a) node-weighted, or (b) directed.
\end{lemma}
We start with (a).
We present a reduction from $\VC$ of the weighted \PS\ problem.
Throughout we consider the special case of the weighted \PS\ problem, where $\Terminals=\{s,t\}$, namely, the \PP\ problem.
In \cite{VCDeg3} it was shown that the $\VC$ problem is NP-complete even for planar graphs with maximum degree $3$. Given a $\VC$ instance in the form of a bounded-degree graph $G(V,E)$, the instance $(G'(V',E',W), s,t)$ to the weighted \PP\ is constructed as follows. We add a ``heavy'' weight neighbor $\widehat{v}$ to every node $v \in V(G)$. In addition, every edge $e \in E$ is replaced by a gadget $g(e)$ consisting of four nodes, $V(g(e))=\{U(e), D(e),L(e),R(e)\}$ forming a diamond graph. Let $S=\bigcup_{e \in E} V(g(e))$ and $\widehat{V}=\bigcup_{v \in V} \{\widehat{v}\}$. The set of nodes of $G'$ is given by $V'=V \cup \widehat{V} \cup S \cup \{s,t\}$. The weight function $W: V \to \mathbb{R}_{>0}$ is defined as follows: $W(v)=1$ if $v \in V' \setminus \widehat{V}$, and $W(v)=|V(G)|+4|E|$ otherwise.

\begin{figure}[h!]
\begin{center}
\includegraphics[scale=0.3]{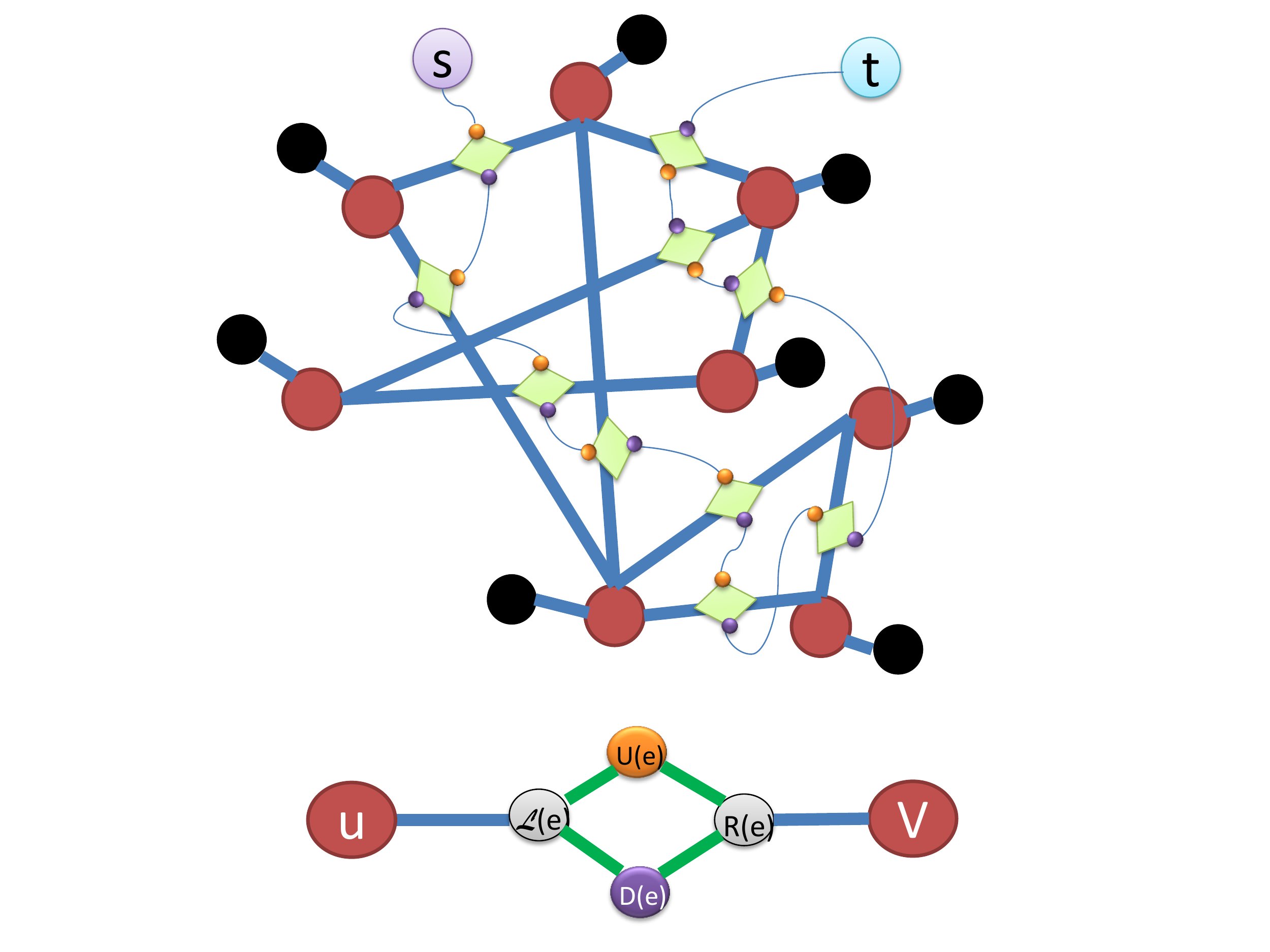}
\hfill
\includegraphics[scale=0.3]{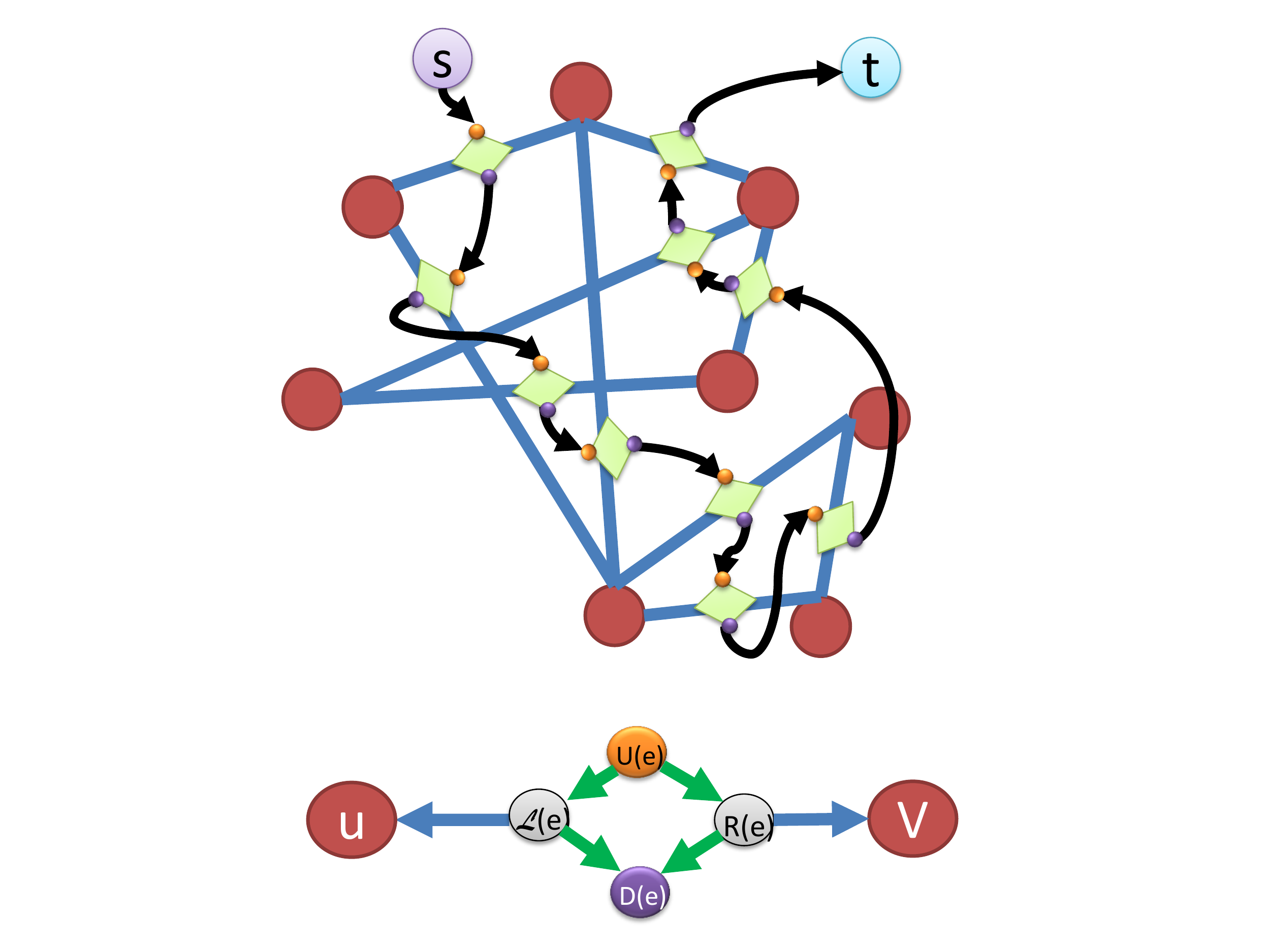}
\caption{ \label{fig:vc}
\sf
Illustration of the reduction from $\VC$. (a) The weighted case. Top: the graph $G$ with the diamond graph gadgets $g(e)$. The black nodes correspond to the heavy neighbor attached to each node $v \in V(G)$. Bottom: zoom into a single gadget. (b) The directed case. Directionality enforces visiting the gadgets in order, precluding the tour in $V(G)$ nodes.
}
\end{center}
\end{figure}

We now describe the gadget construction. Each edge $e=(u,v)$ is replaced by a diamond graph $g(e)$ consisting of the edges $$E(g(e))=\{(u,L(e)), (R(e),v), (L(e), U(e)), (L(e), D(e)), (U(e), R(e)),(D(e), R(e)) \}$$ used to connect $u$ and $v$ in $G'$. (See Fig \ref{fig:vc}(a).) In addition, the new node $s$ (respectively, $t$)
is connected to the node $U(e')$ (respectively, $D(e'')$) for some arbitrary edges $e',e''$, for $e' \neq e''$.
Finally, fix some ordering $e_1, \ldots, e_{m}$ on the edges of $E$ that starts with $e_1=e'$ and ends with $e_m=e''$. Now, for every $i \in \{1, \ldots, m-1\}$ connect $D(e_{i})$ with $U(e_{i+1})$, creating a ``tour'' from $s$ to $t$ via the gadgets. This completes the description of the instance $(G'(V',E',W), s,t)$.
For a path $P$ and set of nodes $W$, denote the set of neighbors of $P$'s nodes in $W$ by
$N_W[P]=N[P] \cap W$.
We now show that $N_{V(G)}[P]$ is an optimal vertex cover for $G$ iff $P$ is
an optimal $s-t$ secluded path in the weighted graph $G'$.
We begin by making an immediate observation.
\begin{observation}
\label{obs:vc_pp}
For every optimal $s-t$ secluded path $P$, $V(P) \cap V(G) =\emptyset$.
\end{observation}
\Proof
Assume the contrary, and let $P$ be an optimal secluded path with at least one node $v \in V(G)$. Since $v$ has a heavy neighbor $\widehat{v}$, it holds that $\Cost(P) \geq 4E+V+1$.
Consider next an alternative path $P'$ that goes strictly through  the nodes in $S$, i.e., follows the tour through the gadgets $g(e)$. Since $N[P'] \subseteq \bigcup_{w \in S} N[w] \subseteq S \cup V(G)$, we get that $\Cost(P) > 4E+V \geq \Cost(P')$, in contradiction to the optimality of $P$.
\QED
Consequently, in the rest of the proof, we consider only secluded paths $P$ that do not visit nodes of $V(G)$. Note that every such $s-t$ path is of length $3|E|$ and the set of $S$ nodes is fully contained in neighbors of this path, i.e., $N_S[P]=S$. It therefore holds that,
\begin{equation}
\label{eq:p_notin_v}
\Cost(P)=|S|+|N_{V(G)}[P]|~.
\end{equation}
The main factor determining the quality of an efficient path, among all paths that do not visit $V(G)$, is the number of path neighbors in $V(G)$, i.e., $|N_{V(G)}[P]|$. That is, by choosing between the $u$-side (respectively, $v$-side) in every gadget $g(e)$, for $e=(u,v)$, the resulting path has $u$ (respectively, $v$) in its neighborhood, i.e., $N_{V(G)}[P]$. Overall, there are $|E|$ such decisions to be made. Let $VC(G)$ be an optimal vertex cover for $G$. We show the following.
\begin{claim}
\label{cl:vc}
$q^{*}_{s,t}=4|E|+|VC(G)|$.
\end{claim}
\Proof
We begin by showing that $\Cost(P)\geq 4|E|+|VC(G)|$ for every $s-t$ path $P$ that does not go through the nodes of $V(G)$.
Let $N^{*}=N_{V(G)}(P)$. We now show that $N^{*}$ is a legal vertex cover of $G$, hence $|VC(G)| \leq |N^{*}|$ and by Eq. (\ref{eq:p_notin_v}), $\Cost(P)\geq 4|E|+|VC(G)|$ as required. Assume for contradiction that there exists an edge $e=(u,v) \in E$ not covered by the nodes of $N^{*}$, i.e., both $u,v \notin N^{*}$. Consider the gadget $g(e)$. Since $P$ does not go through the nodes of $V(G)$, it passes through $U(e)$ and $D(e)$. In particular, when reaching $U(e)$, a decision is made between passing through the $u$ side $L(e)$ or through the $v$ side $R(e)$. Without loss of generality assume that $P$ passes through $L(e)$, i.e., $P=[s, \ldots, U(e),L(e), D(e), \ldots, t]$. Then, since $u$ is a neighbor of $L(e)$, it is in $N^{*}$ and we end with contradiction.

Conversely, we now show that for every legal vertex cover $C$ of $G$, there exists an
$s-t$ secluded path $P$ whose cost satisfies $\Cost(P) \leq |C|+4E$. The path $P$ is constructed as follows. Start from $s$ and traverse the gadgets $g(e)$ in order, from $g(e_1)$ to $g(e_m)$. Within each gadget $g(e)$, for $e=(u,v)$ such that $(u,L(e)) \in E'$ and $(v,R(e)) \in E'$,
move from $U(e)$ to $D(e)$ through $L(e)$ if $u \in C$ and through $R(e)$ otherwise. One of these choices must hold, since $C$ is a legal vertex cover, so it must contain at least one of the nodes $u$ and $v$. Hence, the path $P$ ``pays'' for a node $v$ in $V(G)$ only if $v \in C$. Formally, we get that $N^{*} \subseteq C$. Hence, combining with Eq. (\ref{eq:p_notin_v}), $\Cost(P) \leq 4E+|C|$. The claim follows.
\QED
Part (a) of the lemma follows.
Finally, we now turn consider Lemma \ref{lem:privatepath_vc}(b).
It is sufficient to show a directed version for the reduction such that Obs. \ref{obs:vc_pp} still holds. Note that the main essence is to prevent any optimal path from visiting the nodes of $V(G)$ (and hence to visit every $g(e)$). In the weighted case, this goal is achieved by adding a ``heavy'' unique neighbor to every node in $V(G)$ (the black nodes in Fig. \ref{fig:vc}(a)). In the directed case, this property is achieved by directing the edges $E'$ in the following manner.
Recall the ordering $e_1, \ldots, e_m$ on the edges $E$, which imposes a direction for each edge in our construction as follows.
We include the arcs $(s, U(e_1))$, $(D(e_m), t)$, $(U(e_i), L(e_i))$, $(U(e_i), R(e_i))$, $(L(e_i),D(e_i))$, $(R(e_i),D(e_i))$, and in addition, $(D(e_i), U(e_{i+1}))$ for every $i \in \{1, \ldots,m-1\}$. Finally, the edges between $L(e_i)$ and $R(e_i)$ to the nodes of $V(G)$ are directed towards the nodes of $V(G)$ (see Fig. \ref{fig:vc}(b)).  This forces any $s-t$ path in $G'$ to go through strictly the nodes in $S$. Obs. \ref{obs:vc_pp} now holds for the directed case, and the rest follows as in the weighted case. The lemma follows.
\QED

\subsection{Polynomial-time algorithm for the unweighted undirected case.}
In contrast, for unweighted undirected case of bounded-degree graphs, we show that the \PP\ problem is polynomial.
For two subpaths $P_1,P_2$, define their asymmetric difference as
\begin{equation*}
\DIFF(P_1,P_2)=|N[P_1] \setminus N[P_2]|~.
\end{equation*}
\begin{observation}
\label{obs:delta}
(a) $\DIFF(P_1,P_2) \leq \DIFF(P_1,P')$ for every $P' \subseteq P_2$.\\
(b) $\Cost(P_1 \circ P_2)= \Cost(P_1) + \DIFF(P_2,P_1)$.
\end{observation}
\Proof
Part (a) simply follows as $N[P'] \subseteq N[P_2]$. Part (b) follows by definition.
\QED
We have the following.
\begin{theorem}
\label{thm:bounded_deg_poly}
The \PP\ problem is \emph{polynomial} on unweighted undirected degree-bounded graphs.
\end{theorem}
Note that in the previous section we showed that if the graph is either weighted or directed then the \PP\ problem (i.e., the special case of \PS\ with two terminals) is NP-hard. In addition, it is noteworthy that the related problem of $\MLP$ problem \cite{YuanVJ05, HassinMS07}  is NP-hard even for unweighted planar graph with max degree 4 (this follows from a straightforward reduction from \VC).

We begin with notation and couple of key observations in this context.
For a given path $P$, let $d_{P}(s,u)$ be the distance in edges between $s$ and $u$ in the path $P$. Recall, that $\MAXDEG$ is the maximum degree in graph $G$.
\begin{observation}
\label{obs:opt_path_common}
Let $u,v$ be two nodes in some optimal $s-t$ path $P^{*}$ that share a common neighbor, i.e., $N(u) \cap N(v) \neq \emptyset$. Then $d_{P^{*}}(u,v)\leq \DegAnyConst$.
\end{observation}
\Proof
Assume for contradiction that there exists an optimal $s-t$ path $P^{*}$ such that
$N(u) \cap N(v) \neq \emptyset$ where $u=P^{*}[i]$ and $v=P^{*}[j]$, $1<i <j$ for $(i-j)\geq \MAXDEG+2$.
Recall that due to the optimality of $P^{*}$, for every node $u$ in $P^*$, the only neighbors of $u$ in $G$ from among the nodes of $V(P^{*})$ are the nodes adjacent to $u$ in $P^*$ (otherwise the path can be shortcut). Let $w \in N(u) \cap N(v)$ be the mutual neighbor of $u$ and $v$ and consider the alternative $s-t$ path $\widehat{P}$ obtained from $P^{*}$ by replacing the subpath $Q=P[i, \ldots, j]$ by the subpath $P'=[u, w, v]$. Let $Q^{-}=P[i+2, \ldots, j-2]$ be an length-$\ell'$ internal subpath of $Q$ where $\ell' \geq \MAXDEG-1$. Then since the  degree of $w$ is at most $\MAXDEG$, it follows that
\begin{equation}
\label{eq:path_bounded_new}
\Cost(\widehat{P}) \leq N[V(P)\setminus V(Q^-)]+\MAXDEG-2~,
\end{equation}
where $\MAXDEG-2$ is an upper bound on the number $w's$ neighbors other than $u$ and $v$.
In addition, note that by the optimality of $P^{*}$ it contains no shortcut and hence
$V(Q^{-}) \cap \left(N[V(P)\setminus V(Q^-)] \right)=\emptyset$
(i.e., for node $P[i]$ the only neighbors on the path $P$ are $P[i-1]$ and $P[i+1]$).
Thus,
\begin{eqnarray*}
\Cost(P^{*}) &\geq& N[V(P)\setminus V(Q^-)]+|V(Q^{-})|
\\&\geq&
N[V(P)\setminus V(Q^-)]+\MAXDEG-1 > \Cost(\widehat{P})~,
\end{eqnarray*}
where the last inequality follows from Eq. (\ref{eq:path_bounded_new}), contradicting the optimality of $P^{*}$. The observation follows.
\QED
In other words, the observation says that two vertices on the optimal path at distance $\MAXDEG+2$ (which is constant for bounded-degree graphs) or more have no common neighbors. This key observation is at the heart of our dynamic program, as it enables the necessary subproblem independence property. The difficulty is that this observation applies only to optimal paths, and so a delicate analysis is needed to justify why the dynamic program works.
Note that the main difficulty of computing the optimal secluded path $P^{*}$ is that the cost function $\Cost(P^{*})$ is not a linear function of path's  components as in the related $\DegCost$ measure. Instead, the residual cost of the $i$th vertex in the path \emph{depends} on the neighborhood of the length-$(i-1)$ prefix of $P^{*}$. This dependency implies that the secluded path computation cannot be simply decomposed into independent subtasks.
However, in contrast to suboptimal $s-t$ paths, the dependency (due to mutual neighbors) between the components of an optimal path is \emph{limited} by the maximum degree $\MAXDEG$ of the graph. The \emph{limited dependency} exhibited by any $s-t'$ optimal path facilitates the correctness of the dynamic programming approach. Essentially, in the dynamic program, entries that correspond to subsolution $\sigma$ of \emph{any} optimal $s-t'$ path enjoy the limited dependency, and hence the values computed for these entries correspond to the \emph{exact} cost of an optimal path that starts with $s$ and ends with $\sigma$. In contrast, entries of subpaths $\sigma$ that do not participate in any optimal $s-t'$ path, correspond to an \emph{upper bound} on the cost of some path $P$ that starts with $s$ and ends with $\sigma$. This is due to the fact that the value of the entry is computed under the limited dependency assumption, and thus does not take into account the possible double counting of mutual neighbors between \emph{distant} vertices in the path. Therefore, the possibly ``falsified'' entries cannot compete with the exact values, which are guaranteed to be computed for the entries that hold the subpaths of the optimal path. This informal intuition is formalized below.

For a path $P$ of length $\ell \geq \DegAnyConst$, let $\Suff(P)=\langle v_{\ell-\MAXDEG}, \ldots, v_{\ell} \rangle$ be the $(\DegAnyConst)$-suffix of $P$.
\begin{lemma}
\label{lem:opt_path_common}
Let $P^{*}$ be an optimal $s-t'$ path of length $\geq \DegAnyConst$. Let $P^{*}=P_1 \circ P_2$ be some partition of $P^{*}$ into two subpaths such that $|P_1| \geq \DegAnyConst$. Then $\DIFF(P_2,P_1)=\DIFF(P_2, \Suff(P_1))$.
\end{lemma}
\Proof
For ease of notation, let $\ell=|P_1|$, $P'=P_1[1, \ldots, \ell-(\DegAnyConst)]$, and $\sigma=\Suff(P_1)$. By definition, $\DIFF(P_2,P_1)=|\left(N[P_2] \setminus N[\sigma] \right) \setminus N[P']|$. In the same manner, $\DIFF(P_2,\sigma)=|\left(N[P_2] \setminus N[\sigma] \right)|$.  Assume for contradiction that the lemma does, namely, $\DIFF(P_2,\sigma) \neq \DIFF(P_2,P' \circ \sigma)$. Then by Obs. \ref{obs:delta}(a) we have that $\DIFF(P_2,\sigma)>\DIFF(P_2,P' \circ \sigma)$. This implies that $N[P_2] \cap  N[P']\neq \emptyset$. Let $u \in N[P_2] \cap  N[P']$. There are two cases to consider: (a) $u \in P'$, and (b) $u$ has a neighbor $v_1$ in $P'$. We handle case (a) by further dividing it into two subcases: (a1) $u \in P_2$, i.e., $u$ occurs at least twice in $P^{*}$, once in $P'$ and once in $P_2$, and (a2) $u$ has a neighbor $v_2$ in $P_2$. Note that in both subcases, there exists a shortcut of $P^{*}$, obtained in subcase (a1) by cutting the subpath between the two duplicates of $u$ and in subcase (a2) by shortcutting from $u$ to $v_2$.  This shortcut results in a strictly lower cost path, in contradiction to the optimality of $P^{*}$. We proceed with case (b).
Let $v_2 \in P_2$ be such that $u \in N[v_2]$. If $v_2=u$, then clearly the path can be shortcut by going from $v_1 \in P'$ directly to $u \in P_2$, resulting in a lower cost path, in contradiction to the optimality of $P^{*}$. If $v_2 \neq u$, then $d_{P^{*}}(v_1,v_2)\geq \MAXDEG+2$ (since $|\sigma|\geq \DegAnyConst$) and $N[v_1] \cap N(v_2) \neq \emptyset$, which in contradiction to Obs. \ref{obs:opt_path_common}. The Lemma follows.
\QED
\begin{corollary}
\label{cor:opt_path_common}
Let $P^{*}$ be an optimal $s-t'$ path of length $\geq \DegAnyConst$. Let $P^{*}=P_1 \circ P_2$ be some partition of $P^{*}$ into two subpaths such that $|P_1| \geq \DegAnyConst$. Then $\Cost(P^{*})=\Cost(P_1)+\DIFF(P_2,\Suff(P_1))$.
\end{corollary}
For clarity of representation, we describe a polynomial algorithm for the \PP\ problem, i.e., where $\Terminals=\{s,t\}$ and in addition $\MAXDEG\leq 3$, in which case $\Suff(P)=\langle v_{\ell-3}, \ldots, v_{\ell} \rangle$.
The general case of $\MAXDEG(G)=O(1)$ is immediate by the description for the special case of $\MAXDEG=3$. The case of \PS\ with a fixed number of terminals $|\Terminals|=O(1)$ is described in Subsection \ref{subsec:multiterminals}. 
\par The algorithm we present is based on dynamic programming.
For each $\DegThreeConst \leq i \leq n$, and every length-$\DegThreeConst$ subpath given by a quartet of nodes $\sigma \subseteq V(G)$, it computes a length-$i$ path $\pi(\sigma,i)$ that  starts with $s$ and ends with $\sigma$ (if such exists) and an upper bound $f(\sigma,i)$ on the cost of this path, $f(\sigma,i) \geq \Cost(\pi(\sigma,i))$. These values are computed inductively, using the values previously computed for other $\sigma$'s and $i-1$. In contrast to the general framework of dynamic programming, the interpretation of the computed values $f(\sigma,i)$ and $\pi(\sigma,i)$, namely, the relation between the dynamic programming values $f(\sigma,i)$ and $\pi(\sigma,i)$ and some ``optimal'' counterparts  is more involved. In general, for arbitrary $\sigma$ and $i$, the path $\pi(\sigma,i)$ is not guaranteed to be optimal in any sense and neither is its corresponding value $f(\sigma,i)$ (as $f(\sigma,i) \geq \Cost(\pi(\sigma,i))$).
However, quite interestingly, there is a subset of quartets for which a useful characterization of $f(\sigma,i)$ and $\pi(\sigma,i)$ can be established. Specifically, for every $\DegThreeConst \leq i \leq n$, there is a subclass of quartets $\Psi^{*}_{i}$, for which the computed values are in fact ``optimal'', in the sense that for every $\sigma \in \Psi^{*}_{i}$, the length-$i$ path $\pi(\sigma,i)$ is of minimal cost among all other length-$i$ paths that start with $s$ and end with $\sigma$. We call such a path a \emph{semi-optimal} path, since it is optimal only restricted to the length and specific suffix requirements. It turns out that for the special class of quartets $\Psi^{*}_{i}$, every semi-optimal path of $\sigma \in \Psi^{*}_{i}$ is also a prefix of some optimal $s-t'$ path. This property allows one to apply Cor. \ref{cor:opt_path_common}, which constitutes the key ingredient in our technique. In particular, it allows us to establish that $f(\sigma,i)=\Cost(\pi(\sigma,i))$. This is sufficient for our purposes since the set $\bigcup_{i=1}^{n} \Psi^{*}_{i}$ contains \emph{any} quartet that occurs in \emph{some} optimal secluded $s-t$ path $P^{*}$. Specifically, for every $s-t$ optimal path $P^{*}$, it holds that the quartet $\Suff(P^{*})=P^{*}[i-3,i]$ satisfies $\Suff(P^{*}) \in \Psi^{*}_{i}$. The correctness of the dynamic programming is established by the fact that the values computed for quartets that occur in optimal paths in fact correspond to optimal values as required (despite the fact the these values are ``useless'' for other quartets).

\subsubsection{The algorithm}
If the shortest path between $s$ and $t$ is less than $3$, then the optimal \PP\ can be found by an exhaustive search, so we assume throughout that $d_{G}(s,t)\geq 3$. Let $\Psi$ be the set of all length-\DegThreeConst~ subpaths $\sigma$ in $G$. That is, for every $\sigma\in \Psi$, we have $V(\sigma)\subseteq V(G)$ and $(\sigma[i],\sigma[i+1])\in E(G)$ for every $i \in [1,3]$.
For $\sigma \in \Psi$, define the collection of \emph{shifted successor} of $G$ as $Next(\sigma)=\{ \langle \sigma[2], \ldots, \sigma[\DegThreeConst], u \rangle \mid u \in \left(N[\sigma[\DegThreeConst]] \setminus \sigma \right)\}$. For every pair $(\sigma, i)$, where $\sigma \in \Psi$ and $i \in \{1, \ldots,n\}$, the algorithm computes a value $f(\sigma,i)$ and length-$i$ path $\pi(\sigma,i)$ ending with $\sigma$  (i.e., $\pi(\sigma,i)[i-3, \ldots,i]=\sigma$). These values are computed inductively. For $i=\DegThreeConst$, let $\pi(\sigma,\DegThreeConst)=\sigma$ and
$$f(\sigma,\DegThreeConst) ~=~
\begin{cases}
\Cost(\sigma), & \text{if $\sigma[1]=s$}\\
\infty, & \text{otherwise}
\end{cases}
$$
Once the algorithm has computed $f(\sigma,j)$ for every $\DegThreeConst \leq j\leq i-1$ and every $\sigma \in \Psi$, in step $i \in \{5, n\}$ it computes
\begin{equation}
\label{eq:cost_dynamic}
f(\sigma,i)=\min\{ f(\sigma',i-1)+\DIFF(\sigma,\sigma') ~\mid~ \sigma \in Next(\sigma')\}.
\end{equation}
Note that
\begin{equation}
\label{eq:delta}
\DIFF(\sigma,\sigma')=\DIFF(\sigma[\DegThreeConst],\sigma')=|N[\sigma[\DegThreeConst]] \setminus N[\sigma']|.
\end{equation}
Let $\sigma' \in \Psi$ such that $\sigma \in Next(\sigma')$ and $\sigma'$ achieves the minimum value in  Eq. (\ref{eq:cost_dynamic}). Then define $Pred(\sigma)=\sigma'$ and let
\begin{equation}
\label{eq:path_dynamic}
\pi(\sigma,i)=\pi(Pred(\sigma),i-1) \circ \sigma[\DegThreeConst].
\end{equation}
Let $q^{*}_{s,t}=\min\{f(\sigma,i) \mid i\in \{\DegThreeConst, \ldots, n\}, Last(\pi(\sigma,i))=t\}$, and set $P^{*}=\pi(\sigma^{*},i^{*})$ for $\sigma^{*},i^{*}$ such that $f(\sigma^{*},i^{*})=q^{*}_{s,t}$.
Note that there are at most $O(n^{4} \cdot n)$ entries $f(\sigma,i)$, each computed in constant time, and so the overall running time is $O(n^{5})$.
\vskip .2cm \noindent \textbf{Analysis.}
Throughout this section we make use of two notions of optimality for secluded paths. The two notions are parameterized by $\sigma \in \Psi$ and by an index $i \in \{1, \ldots, n\}$.  Let ${\cal P}(\sigma,i)$ be the set of length-$i$ paths $P$ starting at $P[1]=s$ and ending with $\Suff(P)=\sigma$ for $P \in {\cal P}(\sigma,i)$.
Denote the minimum cost of a path in ${\cal P}(\sigma,i)$ by $q^{*}(\sigma,i)=\min \{\Cost(P) \mid P \in {\cal P}(\sigma,i)\}$
and denote the set of paths attaining this minimum cost by ${\cal Q}_{semi}(\sigma,i)=\{P \in {\cal P}(\sigma,i) \mid \Cost(P)=q^{*}(\sigma,i)\}$.
We refer to paths in ${\cal Q}_{semi}(\sigma,i)$ as \emph{semi-optimal} paths, since they are optimal only among length-$i$ paths ending with $\sigma$.
Note that in general, a semi-optimal path $P \in {\cal Q}_{semi}(\sigma,i)$ is not necessarily an optimal secluded path from $s$ to $P[i]$,
nor is it a prefix of an optimal secluded path from $s$ to any $t'$, and thus Cor. \ref{cor:opt_path_common} cannot be applied.
However, we next argue that there exists a subset of quartets $\sigma^{*} \in \Psi$ for which every path $P \in {\cal Q}_{semi}(\sigma,i)$ is  a prefix of an optimal $s-t'$ secluded path,
and since Obs. \ref{obs:opt_path_common} holds for any subpath of some optimal $s-t'$ path, we can apply it for this subset of sequences.
Let $\mathcal{Q}^{*}_{t'}=\{P \in {\cal P}_{s,t'} \mid \Cost(P)=q^{*}_{s,t'}\}$ be the set of all optimal $s-t'$ paths for  $t' \in V(G) \setminus \{s\}$.
A path $P$ that is a \emph{prefix} of some optimal $s-t'$ path is hereafter referred to as a \emph{pref-optimal} path.
Let $\mathcal{Q}^{*}(\sigma,i)=\{P \in \bigcup_{t' \in V(G) \setminus \{s\}} \mathcal{Q}^{*}_{t'} ~\mid~ P[i-3,i]=\sigma\}$.
That is, $\mathcal{Q}^{*}(\sigma,i)$ is the set of all optimal secluded paths in which the subpath in positions $(i-3)$ to $i$ is equal to $\sigma$. Note that in contrast to the set $\mathcal{Q}_{semi}(\sigma,i)$, which is nonempty for every $\sigma \in \Psi$, and might contain suboptimal  secluded paths, the set  $Q^{*}(\sigma,i)$ might be empty for many choices of $\sigma$, but every path $P \in \mathcal{Q}^{*}(\sigma,i)$ is indeed optimal.
\par Define the subset of subpaths for which $\mathcal{Q}^{*}(\sigma,i)$ is nonempty, for $i\in  \{\DegThreeConst, \ldots,  n\}$, as
$$\Psi^{*}_{i}=\{\sigma \in \Psi ~\mid~ \mathcal{Q}^{*}(\sigma,i) \neq \emptyset\}.$$ We next establish some useful properties for $\sigma \in \Psi^{*}_{i}$. Let $\mathcal{Q}^{T}(\sigma,i)$ correspond to the set of paths in $\mathcal{Q}^{*}(\sigma,i)$ but truncated at position $i$, that is, $\mathcal{Q}^{T}(\sigma,i)=\{P[1,i] \mid P \in \mathcal{Q}^{*}(\sigma,i)\}$. Clearly, every path $P$ in $\mathcal{Q}^{T}(\sigma,i)$ is of length $i$.
Moreover, since $\mathcal{Q}^{*}(\sigma,i)\}$ contains optimal paths, $\mathcal{Q}^{T}(\sigma,i)$ consists of pref-optimal paths.
It is important to understand the distinction between the set $\mathcal{Q}^{T}(\sigma,i)$ and the set $\mathcal{Q}_{semi}(\sigma,i)$. Both sets contain length-$i$ paths that start with $s$ and end with $\sigma$. However, while $P_1 \in \mathcal{Q}^{T}(\sigma,i)$ (if it exists) corresponds to a prefix of some $s-t'$ optimal path, a path $P_2 \in \mathcal{Q}_{semi}(\sigma,i)$ is not necessarily a prefix of some optimal path; it is only optimal among paths of length $i$ ending with $\sigma$. In particular, this difference implies that Cor. \ref{cor:opt_path_common} can be safely applied to paths in $\mathcal{Q}^{T}(\sigma,i)$. \\
\par We first provide a general observation regarding the computed values $f(\sigma,i)$ and $\pi(\sigma,i)$.
\begin{observation}
\label{lem:start}
For every $\sigma \in \Psi$ and $i \in \{1, \ldots, n\}$, if $P^{*} \in \mathcal{Q}_{semi}(\sigma,i)$ then $\Cost(P^{*})\leq \Cost(\pi(\sigma,i)) \leq f(\sigma,i)$.
\end{observation}
\Proof
The left inequality follows immediately, since both $\pi(\sigma,i)$ and $P^{*}$ are in $\mathcal{P}(\sigma,i)$ and $P^{*}$ is a semi-optimal path, i.e., it has the lowest cost among these paths. The right inequality is proved by induction on $i$. for $i=\DegThreeConst$, the claim holds trivially. Assume it holds for every $\sigma$ and every $j \leq i-1$, and consider a fixed $\sigma$ and $i$. Let $\sigma'=Pred(\sigma)$. Then
\begin{eqnarray}
\label{eqn:start}
\Cost(\pi(\sigma,i))&=&\Cost(\pi(\sigma',i-1) \circ \sigma[\DegThreeConst]) \nonumber
\\&=&
\Cost(\pi(\sigma',i-1))+\DIFF(\sigma[\DegThreeConst],\pi(\sigma',i-1)) \label{eq:ineqd}
\\&\leq&
f(\sigma',i-1)+\DIFF(\sigma[\DegThreeConst],\sigma')
\label{eq:ineqsig}
\\&=&
f(\sigma',i-1)+\DIFF(\sigma,\sigma')  \label{eq:ineqsig1}
\\&=&
f(\sigma,i) ~, \label{eq:ineqsig2}
\end{eqnarray}
where Ineq. (\ref{eq:ineqd}) follows from Obs. \ref{obs:delta}(b), Ineq. (\ref{eq:ineqsig}) follows from the inductive assumption, Eq. (\ref{eq:ineqsig1}) follows from Eq. (\ref{eq:delta}) and Eq. (\ref{eqn:start}) follows from Eq. (\ref{eq:cost_dynamic}). The observation follows.
\QED
It turns out that for the set of special quartets that occur on some optimal $s-t$ path, a stronger characterization of $\pi(\sigma,i)$ and $f(\sigma,i)$ can be established.
\begin{lemma}
\label{lem:in_optimal}
For every $i$ and $\sigma \in \Psi^{*}_i$~,\\
(a) $\pi(\sigma,i)\in \mathcal{Q}_{semi}(\sigma,i)$, and \\
(b) $f(\sigma,i)=\Cost(\pi(\sigma,i))$.
\end{lemma}
Let $P^{*}$ be some optimal secluded path and let $\sigma=\Suff(P^{*})$.
Since $\sigma \in \Psi^{*}_{|P^{*}|}$, by Lemma \ref{lem:in_optimal} we get that the computed path $\pi(\sigma,|P^{*}|)$ is indeed an optimal $s-t$ path of cost $f(\sigma,|P^{*}|)$. Our remaining goal is therefore to establish Lemma \ref{lem:in_optimal}.
\vskip .1cm \noindent \textbf{Proof Sketch.}
We prove that the inequalities of Obs. \ref{lem:start} are in fact \emph{equalities} for every $\DegThreeConst \leq i \leq n$ and every $\sigma \in \Psi^{*}_{i}$. The proof, by induction on $i$, is not immediate and requires several properties to be established and to come into play together. To begin, assume that
both parts of  Lemma \ref{lem:in_optimal} hold for every $\sigma \in \Psi^{*}_{i}$ for every $\DegThreeConst \leq i \leq \ell-1$ and consider some $\sigma \in \Psi^{*}_{\ell}$. In addition to the inductive assumption, we assume that $Pred(\sigma) \in \Psi^{*}_{\ell-1}$ (which would be established later on). Given these assumptions, for part (a) of the lemma we show that $\pi(\sigma,\ell)=\pi(Pred(\sigma),\ell-1) \circ \sigma[\DegThreeConst] \in \mathcal{Q}_{semi}(\sigma,\ell)$. Next, to show that part (b)  also holds, i.e., that $f(\sigma,\ell)=\Cost(\pi(\sigma,\ell))$, we would like to show that $\pi(\sigma,\ell)$ is a pref-optimal path and thus Cor. \ref{cor:opt_path_common} can be applied.
Our argumentation can be sketched as follows. We prove that for  every  $\sigma \in \Psi^{*}(\ell)$ it holds that (a) every semi-optimal path $P \in \mathcal{Q}_{semi}(\sigma,\ell)$ is pref-optimal, and (b) there is a set of quartets denoted by $Parent(\sigma,\ell) \subseteq \Psi^{*}_{\ell-1}$, for which the following holds for every $\sigma' \in Parent(\sigma,\ell)$: (b1) $P'=P \circ \sigma[\DegThreeConst] \in \mathcal{Q}_{semi}(\sigma,\ell)$ for every $P \in \mathcal{Q}_{semi}(\sigma',\ell-1)$, and hence $P'$ is pref-optimal, and (b2) $Pred(\sigma) \in Parent(\sigma,\ell)$ and thus the inductive hypothesis can be established.
\paragraph{Detailed analysis.}
We proceed by showing that every semi-optimal length-$i$ path $P \in \mathcal{Q}_{semi}(\sigma,i)$ for $\sigma \in \Psi^{*}_{i}$, $i \in \{\DegThreeConst, \ldots,n\}$ is a pref-optimal path.
\begin{lemma}
\label{lem:con_subopt}
For every $\sigma \in \Psi^{*}_{i}$, $\mathcal{Q}^{T}(\sigma,i)=\mathcal{Q}_{semi}(\sigma,i)$.
\end{lemma}
\Proof
We first show that $\mathcal{Q}^{T}(\sigma,i) \subseteq \mathcal{Q}_{semi}(\sigma,i)$. To do that, we consider a path $P_1 \in \mathcal{Q}^{*}(\sigma,i)$ for which $P_1[1,i] \in \mathcal{Q}^{T}(\sigma,i)$,
and show that $\Cost(P_1[1, i])\leq q^{*}(\sigma,i)$. Thus $P_1[1,i] \in \mathcal{Q}_{semi}(\sigma,i)$.
Let $\ell=|P_1|$ and $t'=\Last(P_1)$, thus $P_1$ is an optimal $s-t'$ path.
Assume for contradiction that $\Cost(P_1[1, i])> q^{*}(\sigma,i)$. Let $P_2 \in \mathcal{Q}_{semi}(\sigma,i)$, i.e., $\Cost(P_2)=q^{*}(\sigma,i)$.  Consider an alternative $s-t'$ path $P_3=P_2 \circ P_1[i+1,\ell]$. Note that $P_3$ is indeed a legal $s-t'$ path since $P_2[i]=P_1[i]$. We now compute its cost and compare it to that of $P_1$.
\begin{eqnarray}
\Cost(P_3)&=& \Cost(P_2) + \DIFF(P_1[i+1,\ell],P_2) \label{ineq:0}
\\&\leq&
\Cost(P_2) + \DIFF(P_1[i+1,\ell],\sigma) \label{ineq:1}
\\&<&
\Cost(P_1[1, i]) + \DIFF(P_1[i+1,\ell],\sigma) \label{ineq:2}
\\&=&
\Cost(P_1), \label{ineq:3}
\end{eqnarray}
where Ineq. (\ref{ineq:0}) and (\ref{ineq:1}) follow by parts (b) and (a) of Obs. \ref{obs:delta} and Ineq. (\ref{ineq:2}) is by the contradiction assumption. Eq. (\ref{ineq:3}) follows from the fact that $P_1$ is an optimal $s-t'$ path and thus Cor. \ref{cor:opt_path_common} can be safely applied.
We therefore get that $\Cost(P_3)<\Cost(P_1)$, in contradiction to the optimality of $P_1$.
\par Conversely, we now prove that $\mathcal{Q}_{semi}(\sigma,i) \subseteq \mathcal{Q}^{T}(\sigma,i)$.
Let $P_1 \in \mathcal{Q}_{semi}(\sigma,i)$ and $P_2 \in \mathcal{Q}^{*}(\sigma,i)$. Let $\ell=|P_2|$ and $t'=\Last(P_2)$. Consider the path $P_3=P_1 \circ P_2[i+1, \ell]$ obtained by replacing the length-$i$ prefix of $P_2$ with $P_1$. We show that $P_3$ has the same cost, $\Cost(P_3)=\Cost(P_2)$, hence $P_3$ is also an optimal  $s-t'$ path. This would imply that $P_3 \in \mathcal{Q}^{*}(\sigma,i)$ and $P_1 \in \mathcal{Q}^{T}(\sigma,i)$, establishing the lemma.
\begin{eqnarray}
\Cost(P_3)&=& \Cost(P_1 \circ P_2[i+1, \ell])= \Cost(P_1) +\DIFF(P_2[i+1, \ell], P_1) \nonumber
\\& \leq &
\Cost(P_1) +\DIFF(P_2[i+1, \ell], \sigma) \nonumber
\\& \leq &
\Cost(P_2[1,i]) +\DIFF(P_2[i+1, \ell], \sigma) \label{eq:con1}
\\& = &
\Cost(P_2)~, \label{eq:con2}
\end{eqnarray}
where Ineq. (\ref{eq:con1}) follows from the fact that both $P_1$ and $P_2[1,i]$ are in $\mathcal{P}(\sigma,i)$, i.e., both are length-$i$ paths starting at $s$ and ending with $\sigma$ and $P_1$ is semi-optimal. Ineq. (\ref{eq:con2}) follows from the fact that $P_2$ is an optimal $s-t'$ path, thus Cor. \ref{cor:opt_path_common} can be safely applied. The fact that $P_2$ is optimal necessitates equality. The lemma follows.
\QED
For $\sigma \in \Psi^{*}_{i}$ and $i \in \{5, \ldots, n\}$, define $$Parent(\sigma,i)=\{\sigma'= \langle u,\sigma[1], \ldots, \sigma[3] \rangle~\mid~ \exists P \in \mathcal{Q}^{*}(\sigma,i), P[i-\DegThreeConst,i]=u \circ \sigma\}.$$
That is, $\sigma' \in Parent(\sigma,i)$ iff there exist some $t' \in V(G) \setminus \{s\}$ and an optimal $s-t'$ path $P^{*}$ such that $P^{*}[i-\DegThreeConst,i]=\sigma'(1) \circ \sigma$.
Note that by the definition of $\Psi^{*}_{i}$, there exists at least one pref-optimal path $P$ of length $i$ that ends with $\sigma$. In addition, since $i\geq 5$ there exists at least one $\sigma' \in Parent(\sigma,i)$ corresponding to $\sigma'=P[i-\DegThreeConst, i-1]$. In other words, $Parent(\sigma,i) \neq \emptyset$ for every $i\geq 5$ and $\sigma \in \Psi^{*}_{i}$.
Lemma \ref{lem:con_subopt} implies that the following holds for every $\sigma \in \Psi_{i}^{*}$.
\begin{observation}
\label{cl:parent_opt}
If $\sigma \in \Psi^{*}_{i}$ then \\
(a) for every $\sigma' \in Parent(\sigma,i)$ there exists a semi-optimal path $P^{*} \in \mathcal{Q}_{semi}(\sigma,i)$, which is also a pref-optimal path, such that $P^{*}[i-\DegThreeConst,i]=\sigma' \circ \sigma[\DegThreeConst]$.\\
(b) $Parent(\sigma,i) \subseteq \Psi^{*}_{i-1}$.
\end{observation}
%
We first provide an auxiliary claim that holds for every $i \in \{1, \ldots, n\}$. Let $\sigma,\sigma'$ be such that (a) $\sigma \in \Psi^{*}_{i}$ and
(b) $\sigma' \in Parent(\sigma,i)$.
\begin{claim}
\label{cl:nn}
For every $\widehat{P} \in \mathcal{Q}_{semi}(\sigma',i-1)$, $\widehat{P} \circ \sigma[\DegThreeConst] \in  \mathcal{Q}_{semi}(\sigma,i)$
\end{claim}
\Proof
Let $P \in \mathcal{Q}_{semi}(\sigma,i)$ such that $\sigma'=P[i-\DegThreeConst,i-1]$. Note that since $\sigma' \in Parent(\sigma,i)$, by Obs. \ref{cl:parent_opt}(a), such a $P$ is guaranteed to exist. Let $\widehat{P} \in \mathcal{Q}_{semi}(\sigma',i-1)$ and let $P''=\widehat{P} \circ \sigma[\DegThreeConst]$.
We now prove that $\Cost(P'') \leq \Cost(P)$ and therefore $P'' \in \mathcal{Q}_{semi}(\sigma,i)$.
\begin{eqnarray}
\Cost(P'')&=& \Cost(\widehat{P} \circ \sigma[\DegThreeConst]) =
\Cost(\widehat{P})+\DIFF(\sigma[\DegThreeConst],\widehat{P}) \nonumber
\\&\leq&
\Cost(\widehat{P})+
\DIFF(\sigma[\DegThreeConst],\sigma') \label{eq:sig0}
\\&\leq&
\Cost(P[1,i-1])+\DIFF(\sigma[\DegThreeConst],\sigma')\label{eq:sig1}
\\&=&
\Cost(P) ~, \label{eq:sig2}
\end{eqnarray}
where Ineq. (\ref{eq:sig0}) follows from Obs. \ref{obs:delta}(a), Ineq. (\ref{eq:sig1}) follows from the fact that both $P[1,i-1],\widehat{P} \in \mathcal{P}(\sigma',i-1)$ and $\widehat{P}$, being semi-optimal, has the minimal cost in this set. Eq. (\ref{eq:sig2}) follows from the fact that $P \in \mathcal{Q}_{semi}(\sigma,i)$ and since $\sigma \in \Psi^{*}_i$, Lemma \ref{lem:con_subopt} implies that $P$ is a pref-optimal, i.e., a prefix of some optimal secluded path, i.e., $P \in \mathcal{Q}^{T}(\sigma,i)$. Therefore, Cor. \ref{cor:opt_path_common} can be applied.
\QED
We now turn to proving Lemma \ref{lem:in_optimal}, showing that for every $i$, and every $\sigma \in \Psi^{*}_i$, the inequalities of Obs. \ref{lem:start} become equalities.
\inline Proof of Lemma \ref{lem:in_optimal}:
The lemma is proven by induction on $i$. For $i=\DegThreeConst$, both claims clearly hold. Assume the claims hold for $i \leq \ell-1$ and consider $\ell$. Let $\sigma'=Pred(\sigma)$. To be able to apply the inductive hypothesis to $\sigma'$, we first show the following.
\begin{claim}
\label{cl:parent}
$\sigma' \in Parent(\sigma,\ell)$.
\end{claim}
\Proof
Note that since $\sigma \in \Psi^{*}_{\ell}$, it holds that $Parent(\sigma,\ell) \neq \emptyset$, and also $Parent(\sigma,\ell) \subseteq \Psi^{*}_{\ell-1}$ by Obs. \ref{cl:parent_opt}(b).
Let $\sigma^{*} \in Parent(\sigma,\ell)$, thus $\sigma^{*} \in \Psi^{*}_{\ell-1}$.
By Lemma \ref{lem:con_subopt}, $\mathcal{Q}^{T}(\sigma,\ell)=\mathcal{Q}_{semi}(\sigma,\ell)$  and $\mathcal{Q}^{T}(\sigma^{*},\ell-1)=\mathcal{Q}_{semi}(\sigma^{*},\ell-1)$.
Let us pick three arbitrary representatives $P^{*}(\sigma^{*},\ell-1) \in \mathcal{Q}_{semi}(\sigma^{*},\ell-1)$, $P^{*}(\sigma',\ell-1) \in \mathcal{Q}_{semi}(\sigma',\ell-1)$, $P^{*}(\sigma,\ell) \in \mathcal{Q}_{semi}(\sigma,\ell)$. Recall that due to Lemma \ref{lem:con_subopt}, $P^{*}(\sigma^{*},\ell-1)$ and $P^{*}(\sigma,\ell)$ are pref-optimal paths and therefore Cor. \ref{cor:opt_path_common} can be safely applied for them. We first show that \begin{equation}
\label{eq:parent_opt}
\Cost(P^{*}(\sigma,\ell))=\Cost(P^{*}(\sigma^{*},\ell-1))+\DIFF(\sigma,\sigma^{*}).
\end{equation}
To see this, let $P'=P^{*}(\sigma^{*},\ell-1) \circ \sigma[\DegThreeConst]$. Then by Claim \ref{cl:nn}, $P' \in \mathcal{Q}_{semi}(\sigma,\ell)$. In addition, by Lemma \ref{lem:con_subopt}, $P' \in \mathcal{Q}^{T}(\sigma,\ell)$ and therefore Cor. \ref{cor:opt_path_common} can be safely applied, so Eq. (\ref{eq:parent_opt}) is established. We next show that $P''=P^{*}(\sigma',\ell-1) \circ \sigma[\DegThreeConst] \in \mathcal{Q}_{semi}(\sigma,\ell)$.  Clearly, $P'' \in \mathcal{P}(\sigma,\ell)$, so it suffices to show that $\Cost(P'') \leq \Cost(P^{*}(\sigma,\ell))$ as $P^{*}(\sigma,\ell)$ is semi-optimal. Indeed,
\begin{eqnarray}
\Cost(P'') &=& \Cost(P^{*}(\sigma',\ell-1))+\DIFF(\sigma[\DegThreeConst],P^{*}(\sigma',\ell-1)) \nonumber
\\&\leq&
\Cost(P^{*}(\sigma',\ell-1))+\DIFF(\sigma[\DegThreeConst],\sigma') \label{eq:ineq4}
\\&\leq&
f(\sigma',\ell-1)+\DIFF(\sigma,\sigma') \label{eq:ineq5}
\\&\leq &
f(\sigma^{*},\ell-1)+\DIFF(\sigma,\sigma^{*}) \label{eq:ineq6}
\\&=&
\Cost(\pi(\sigma^{*},\ell-1))+\DIFF(\sigma,\sigma^{*}) \label{eq:ineq7}
\\&= &
\Cost(P^{*}(\sigma^{*},\ell-1))+\DIFF(\sigma,\sigma^{*})= \Cost(P^{*}(\sigma,\ell)) \label{eq:ineq8}
\end{eqnarray}
where Ineq. (\ref{eq:ineq4}) follows from Obs. \ref{obs:delta}(a) and Ineq. (\ref{eq:ineq5}) follows from Lemma \ref{lem:start} and Eq. (\ref{eq:delta}). Ineq. (\ref{eq:ineq6}) follows from the fact that $\sigma \in Next(\sigma')\cap Next(\sigma^{*})$ and $\sigma'$ attains the minimum value of Eq. (\ref{eq:cost_dynamic}). Eq. (\ref{eq:ineq7}) follows from part (b) of the inductive assumption for $\ell-1$. In Eq. (\ref{eq:ineq8}) the first equality follows from the fact that both $\pi(\sigma^{*},\ell-1)$ and $P^{*}(\sigma^{*},\ell-1)$ are in $\mathcal{Q}_{semi}(\sigma^{*},\ell-1)$, and thus $\Cost(P^{*}(\sigma^{*},\ell-1))=\Cost(\pi(\sigma^{*},\ell-1))=q^{*}(\sigma^{*},\ell-1)$, and the second equality follows from Eq.  (\ref{eq:parent_opt}).
\par We therefore get that $P'' \in \mathcal{Q}_{semi}(\sigma,\ell)$ and since $\sigma \in \Psi^{*}_{\ell}$, by Lemma \ref{lem:con_subopt} it holds that $P''\in \mathcal{Q}^{T}(\sigma,\ell)$, i.e., $P''=P^{*}(\sigma',\ell-1) \circ \sigma[\DegThreeConst]$ is also a pref-optimal path. This implies that $\sigma' \in Parent(\sigma,\ell)$, as required.
\QED
By Claim \ref{cl:parent} and Obs. \ref{cl:parent_opt}(b), $\sigma' \in \Psi^{*}_{\ell-1}$. It follows that the inductive hypothesis can be applied to $\sigma'$. By part (a) of the inductive assumption for $\ell-1$, it holds that $\pi(\sigma',\ell-1) \in \mathcal{Q}_{semi}(\sigma',\ell-1)$. Finally, by Claim \ref{cl:nn} we get that the path $\pi(\sigma,\ell)$ chosen by the algorithm satisfies $\pi(\sigma,\ell)=\pi(\sigma',\ell-1)\circ \sigma[\DegThreeConst] \in \mathcal{Q}_{semi}(\sigma,\ell)$, which establishes (a).
We now consider (b). Note that
\begin{eqnarray}
f(\sigma,\ell)&=&f(\sigma',\ell-1)+\DIFF(\sigma,\sigma') \nonumber
\\&=&
\Cost(\pi(\sigma',\ell-1))+\DIFF(\sigma,\sigma') \label{eq:eqeq1}
\\&=&
\Cost(\pi(\sigma',\ell-1) \circ \sigma[\DegThreeConst]) \label{eq:eqeq2}~,
\end{eqnarray}
where Eq. (\ref{eq:eqeq1}) follows from the inductive hypothesis for $\ell-1$ and Eq. (\ref{eq:eqeq2}) follows from the fact that $\pi(\sigma',\ell-1) \circ \sigma[\DegThreeConst] \in \mathcal{Q}^{T}(\sigma,\ell)$ and thus Cor. \ref{cor:opt_path_common} can be applied.
This completes the proof of Lemma \ref{lem:in_optimal}.
We are now ready to complete the correctness proof of the algorithm. Let $P^{*}$ be an optimal $s-t$ path of length $\ell$. Then $P^{*}[\ell-3,\ell] \in \Psi^{*}_{\ell}$. Therefore, by Lemma \ref{lem:in_optimal}, $P^{*}[1,\ell] \in \mathcal{Q}_{semi}(P^{*}[\ell-3,\ell],\ell)$, and so $f(P^{*}[\ell-3,\ell],\ell)=\Cost(P^{*})$, and by Lemma \ref{lem:start}, the cost of any other $\sigma$ such that $t \in \sigma$, and every $i$, satisfies $f(\sigma,i)\geq \pi(\sigma,i) \geq t$.  Thm. \ref{thm:bounded_deg_poly} is established.
\begin{corollary}
\label{cor:deg_pp}
The \PP\ problem can be solved in $O(n^{\MAXDEG+3})$ time.
\end{corollary}
This follows since there are $O(n^{\MAXDEG+1})$ distinct length-($\MAXDEG+1$) subpaths, and the length of the path is bounded by $n$. Hence, overall there are $O(n^{\MAXDEG+2})$ entries and the value for each entry is computed in $O(\MAXDEG)$ time.
For the \PS\ problem, we show the following.
\begin{theorem}
\label{thm:steiner_bounded_deg_poly}
On unweighted undirected degree-bounded graphs, we have the following:
(a) for arbitrary $k$, the \PS\ problem is NP-hard; (b) for $k=O(1)$, the \PS\ problem is \emph{polynomial}.
\end{theorem}
\Proof
We start with (a).
We prove this by reduction from \VC. By \cite{VCDeg3}, we know that \VC~ is NP-complete even for planar graphs with maximum degree $3$.
Given an instance of the \VC~ problem consists of a bounded-degree graph $G=(V,E)$, we transform it into an instance of  the \PS\ problem that consists of a bounded-degree graph $\widehat{G}$ and a set of terminal $\Terminals$. Let $\widehat{B}$ be a complete binary tree with $n$ leaves $L=\{\ell_1, \ldots, \ell_n\}$.
Consider some ordering on $V(G)$ and connect $\ell_i$ to $v_i$, for every $i \in \{1, \ldots, n\}$. In addition, for each vertex $v_i \in V(G)$, add a single neighbor $\widehat{v}_i$ in $\widehat{G}$. Next, for each edge $e=(v_i,v_j) \in E(G)$, add a vertex $v'_{e}$. Every edge $e \in E(G)$ is replaced by the two edges $(v'_{e},v_i)$ and $(v'_{e},v_j)$ in $\widehat{G}$. In sum, the graph $\widehat{G}$ is defined by $$V(\widehat{G})=V(G) \cup V(\widehat{B}) \cup \bigcup_{v_i\in V(G)}\{\widehat{v}_i\} \cup \bigcup_{e \in E(G)}\{v'_{e}\},$$ and  $$E(\widehat{G})=E(\widehat{B}) \cup \bigcup_{i=1}^{n} \{(\ell_i, v_i), (\widehat{v}_i, v_i)\} \cup \bigcup_{e \in E(G)}\{(v'_{e},v_i),(v'_{e},v_j)\}.$$
Note that the maximum degree $\MAXDEG(\widehat{G})$ in $\widehat{G}$ is at most $\MAXDEG(G)+1$. Hence $\widehat{G}$ is a bounded-degree graph as well.
The terminal set $\Terminals=\bigcup_{e \in E(G)}\{v'_{e}\} \cup V(\widehat{B})$ includes the $v'_{e}$ nodes and the nodes of the binary tree $V(\widehat{B})$. We now show that $G$ has vertex cover of size $k$ iff there exists a secluded Steiner tree $T$ in $\widehat{G}$ of cost $|\Terminals|+|V(G)|+k$. Recall that $VC(G)$ is some optimal vertex cover in $G$ and $q^{*}(\Terminals)$ is the cost of the optimal secluded Steiner tree in $\widehat{G}$.
We now show that
\begin{equation*}
q^{*}(\Terminals)=|\Terminals|+|V(G)|+|VC(G)|~.
\end{equation*}
We begin by showing that $q^{*}(\Terminals)\leq |\Terminals|+|V(G)|+|VC(G)|$.
Given a vertex cover $C'$ for $G$, we show that there exists a Steiner tree $T \subseteq \widehat{G}$ of cost $\Cost(T) \leq |\Terminals|+|V(G)|+|C'|$.
The tree $T$ is given by $V(T)=\Terminals \cup C'$
and $E(T)=E(\widehat{B}) \cup \{(\ell_i, v_i) \mid v_i \in C'\} \cup \{ (v'_{e},v_i), e=(v_i,v_j) \in E(G), v_i \in VC(G)\}$.
Since $C'$ is a legal vertex cover in $G$, it follows that every terminal $v'_{e}$ has at least one neighbor $v_i$ (corresponding to the endpoints of $e$ in $G$) that connects it to the tree $T$. Hence, $T$ is a valid Steiner tree. We now analyze its cost. The cost of $T$ consists of the set $V_1=\Terminals \cup V(G)$ of the terminals and their neighbors, and the set $V_2=\{\widehat{v}_i \mid v_i \in C'\}$. Therefore, $\Cost(T)=|V_1|+|V_2|=|\Terminals| + |V(G)|+|C'|$ and
thus $q^{*}(\Terminals) \leq |\Terminals| + |V(G)|+|VC(G)|$ as required.

It remains to show that $|VC(G)| \leq q^{*}(\Terminals)-(|\Terminals|+|V(G)|)$.
Given a Steiner tree $T$ in $\widehat{G}$, we show how to obtain  a legal vertex cover $C$ such that $|C| \leq \Cost(T)-(|\Terminals|+|V(G)|)$.
Let $C=V(T) \cap V(G)$. First, note that $C$ is a vertex cover. Assume for contradiction that it is not, and let $e=(v_i,v_j)$ be a non-covered edge by $C$.
Since neither $v_i$ nor $v_j$ in $C$, we get that $v'_{e}$ is disconnected in $T$, in contradiction to the fact that $T$ is a Steiner tree. Finally, observe that
$\Cost(T)=|\Terminals|+|V(G)|+|V(G) \cap V(T)|$. Since $C=|V(G) \cap V(T)|$,
part (a) of the theorem is established.
We now turn to prove Thm. \ref{thm:bounded_deg_poly}(b) and show that there exists a polynomial-time algorithm if the number of terminals is bounded by a constant.
\subsection{Polynomial-time algorithm for \PS\ with $|\Terminals|=O(1)$.}
\label{subsec:multiterminals}
For simplicity we present the algorithm for maximum degree three (and note that the proof works for any fixed bounded degree).
Let $k=|\Terminals|$. We note that the main difference with \PS\ problem is that here we also need to take into account the ``skeleton'' of the tree.
We iterate over possible skeletons of trees, where by skeleton of a tree we mean the following.
Consider some tree $T$ and contract every path $P \in \Sigma(T)$ such that all nodes in $P$ but the endpoints are of degree 2 and are not terminal into a single edge. The resulting graph is the skeleton of the tree $T$. Note that it is enough to look on trees $T$ such that all leaves are terminals (as otherwise we can take a subtree of the tree spanning all terminals).
Note that the skeleton may contain at most $k$ nodes that are not terminals.
Given the (at most $k$) nodes that are not terminals, an upper bound on the number of different skeletons is the number of different trees with at most $2k$ nodes, which is $O((2k)^{2k - 2})$. There are $O(n^{k})$ options to choose the nonterminal nodes in the skeleton.
We thus get that there are at most $O(n^{k} (2k)^{2k - 2})$ different skeletons.

A subtree $S_{+}$ is an extended skeleton of the skeleton $S$ if it obtained by replacing each edge $e_i=(x_i,y_i)$ in the skeleton $S$ with a path $P_i$ of the following form. Either the path $P_i$ is a path of $G$ of length at most 9 (and contains at most 8 internal nodes), or the prefix of the path $P_i$ is a subpath of length 4 in $G$ from $x_i$ to some node $z_i$ and the suffix of $P_i$ is a subpath of length 4 in $G$ from some node $w_i$ to $y_i$, the nodes $z_i$ and $w_i$ are connected by an imaginary edge (which will be replaced later by a path from $G$).

The algorithm for finding the tree with the lowest cost  is as follows.
Iterate over all possible skeletons $S$ and for each skeleton $S$ iterate over all possible extended skeleton $S_{+}$.
For each $S_{+}$ do the following. Let $V(S_{+})$ be the set of nodes in $S_{+}$.
For each imaginary edge $(x,y)$ in $S_{+}$, consider the graph $G'(x,y)$ that is obtained by removing all nodes $V(S_{+}) \setminus \{x,y\}$ from $G$.
Find the lowest cost \PP\ $\hat{P}(x,y)$ from $x$ to $y$ in $G'(x,y)$.
For each imaginary edge $(x,y)$ in $S_{+}$, replace the edge with the path $\hat{P}(x,y)$. Let $T(S_{+})$ be the tree obtained through this process.
Finally, return the tree $T(\hat{S}_{+})$ with the lowest cost.
Let $T_{alg}$ be the returned tree.

We call nodes in a tree $T$ splitting nodes if their degree in $T$ is greater than 2.
We call a node special in $T$ if it is either a splitting node or a terminal.
Consider the optimal tree $T^*$ and let $S^*$ be its skeleton and $S^{*}_{+}$ be its extended skeleton.
We claim that $\Cost(T(S^{*}_{+})) = \Cost(T^{*})$, this will imply the correctness of the algorithm.

As in the \PP\ problem we may restrict ourselves only to trees such that all nodes that their parent are not special do not have edges to nodes on the tree that are not in their subtree, similarly, every node that none of whose children is special has no edges to nodes in its subtree (otherwise we can find a different tree with a subset of the nodes of $T^*$). In other words, all nodes in $T^*$ that are neither special nor have a neighbor in $T$ that is special have no edges in $G$ to other nodes in $T^*$ that are not adjacent to them in $T^*$.

The following observation is an extension of Obs. \ref {obs:opt_path_common} and is crucial for the algorithm.
\begin{observation}
\label{obs:tree-neighbors}
Consider an edge $e_1$ in the skeleton $S^*$ and their corresponding path $P_1=(z_1,...,z_{r_1}) \in \Sigma(T^{*})$.
The node $z_i$ for $6 \leq i \leq  r_1-6$ cannot have a common neighbor with any other node $z \in T^* \setminus P_1$.
\end{observation}



\begin{lemma}
$\Cost(T(S^{*}_{+})) = \Cost(T^{*})$
\end{lemma}
\Proof
The proof follows from Obs. \ref{obs:tree-neighbors}.
Note that in the optimal solution $T^*$, $N[S^{*}_{+}]\subseteq  N[T^*]$.
Consider an imaginary edge $(x,y)$ in $S^{*}_{+}$. Let $P^*(x,y)$ be the subpath from $x$ to $y$ in $T^*$.
Note that the subpath $P^*(x',y')$ are disjoint for all imaginary edges $(x',y')$ in $S^{*}_{+}$ .
Let $c^*(x',y') = |N[P^*(x',y')] \setminus N[S^{*}_{+}]|$.
Note that $\Cost(T^*) = \sum_{(x',y')}{c^*(x',y')} + |N[S^{*}_{+}]|$.
Note also that $|N[\hat{P}(x,y),G'(x,y)]| \leq  c^*(x',y')$.
We get that $\Cost(T(S^{*}_{+})) \leq \sum_{(x,y)}{|N[\hat{P}(x,y),G'(x,y)]|} +  |N[S^{*}_{+}]| \leq \Cost(T^{*})$.
\QED

\section{Secluded Connectivity for Specific Graph Families}
\vskip .1cm \noindent \textbf{Bounded-treewidth graphs.}
For a graph $G(V,E)$, let $TW(G)$ denote the \emph{treewidth} of $G$.
In this section, we consider graphs $G$ of constant treewidth, i.e. $TW(G)=O(1)$. A \emph{separation} of a graph $G=(V,E)$ is a triplet $(A,B,S)$ where $A \cup B \cup S=V$, $A \cap B \subset S$ and $(u,v) \notin E$ for every $u \in A \setminus S$ and $v \in B \setminus S$. The set $S$ then separates $A \setminus S$ and $B \setminus S$ in $G$.
The concept of treewidth was introduced by Robertson and Seymour \cite{RobertsonS86} using tree-decompositions. (See \cite{TWBodlaender93, Bodlaender07} for an in-depth introduction to this topic.)
For a graph $G=(V,E)$, a tree decomposition is a pair $(\TD,\mathfrak{X})$ consisting of a tree $\TD=(I,A)$ and a collection $\mathfrak{X}=\{X_i\}_{i \in I}$ of vertex subsets (called \emph{bags}) with the following properties:
(T1) Every vertex $v \in V(G)$ is contained in the least one bag $X_i$. For every edge $(u,v)\in E$ there is at least one bag $X_i$ containing both vertices $u,v$. (T2) For every vertex $v \in V(G)$, the nodes $i$ of $\TD$ with $v \in X_i$ form a subtree of $\TD$.
\par To avoid confusion, the elements of $I$ are referred to as \emph{nodes} and the elements of $V(G)$ are referred to as vertices. In addition, we may informally refer to the node $X_i$ by its index $i$. Let $X_i^{+}$ be the set containing the vertices of node $i$ itself, and the vertices in the bags of its descendants in $\TD$. Let $U_{i}$ (respectively, $U_{i}^{+}$) be the set of terminals in $X_i$ (respectively, $X_i^{+}$).
Let $\omega$ be some upper bound on $TW(G)$ such that $|X_i| \leq \omega-1$ for every $i \in I$. Let $U_i^{+}$ be the set of terminals in $X_i^{+}$.
We show the following.
\begin{theorem}
\label{thm:tw_correctness}
The \PS\ problem (and hence also the \PP~problem) can be solved in \emph{linear} time for graphs with fixed treewidth. In addition, given the tree decomposition of $G$, the \PS\ problem is solvable in $\widetilde{O}(n^3)$ if $TW(G)=O(\log n/ \log\log n)$. This holds even for weighted and directed graphs.
\end{theorem}
If the tree decomposition is not known, then for fixed treewidth graphs it can be computed in linear time \cite{Bodlaender93}.
For simplicity, we consider the unweighted undirected case. The weighted and directed cases are considered by the end of the analysis. Given a tree $T$ in $G$ and a separation $(A,B,S)$ in $G$, let $F_A=T(A)$ (respectively, $F_B=T(B)$) be the forest obtained by considering the edges of $T$ restricted to $A$ (respectively, $B$) nodes. Let $\DIFF(A,S)=N[T] \cap (A \setminus S)$
and $\DIFF(B,S)=N[T] \cap (B \setminus S)$. Finally, let $\Cost(T,S)=N[T] \cap S$.
Since $(A,B,S)$ is a separation, it follows that
\begin{equation}
\label{eq:treewidth_cost}
\Cost(T)=\DIFF(A,S)+\DIFF(B,S)+\Cost(T,S).
\end{equation}
Such decomposition of the cost function $\Cost(G')$ holds for any connectivity construct $G' \subseteq G$. Indeed, the correctness of the dynamic programming applied on the tree-decomposition of $G$ is established due to this cost independency property of Eq. (\ref{eq:treewidth_cost}).
\subsubsection{Algorithm}
The \PS\ algorithm is an extension of the algorithm of \cite{ChimaniMZ11} for computing an optimal $\Steiner$ in a bounded-treewidth graph.
The input to the algorithm is a connected graph $G=(V,E)$, a set of terminals $\Terminals$ and a tree decomposition $\TD=(I,A)$ of $G$.
For ease of analysis, we assume that the tree decomposition $\TD$ is \emph{nice} in the sense that every node $i \in I$ has at most two children $i_1, i_2 \in I$ in the tree decomposition $\TD$ (as in \cite{ChimaniMZ11}).
\par Initially choose any root node $r \in I$ that  contains at least one terminal, i.e. $X_r \cap \Terminals \neq \emptyset$, and direct the tree $\TD$ such that $r$ is the root.
For each $i \in I$ we maintain a table $tb(i)$. Each entry $\kappa_i \in tb(i)$ in the table represents a configuration, implicitly corresponding to a subsolution. Such a subsolution is a forest $F \subseteq G$ that when imposed on the nodes of $X_i^{+}$, resulting in $F(X_i^{+})$, spans the terminals $U_i^{+}$ in $X_i^{+}$, and $F(X_i^{+})$ may be extended into a tree that spans all the terminals. Specifically, the configurations define a collection of disjoint subsets of the vertices of $X_i$ and a collection of edges between vertices in $X_i$. In addition, it specifies the vertices $\widetilde{v} \in X_i$ that are neighbors of the forest $F$ (note that for such $\widetilde{v}$, its neighbor in $F$ is not necessarily in $X_i$; it might be in an ancestor of $X_i$ in $\TD$) . We follow the representation scheme of \cite{ChimaniMZ11} where each subsolution is represented by a color assignment $\kappa_i$ to the vertices of $X_i$, and extend it by adding an additional color $\omega$ that represents the status of being a neighbor. We refer to this coloring as a \emph{configuration}. The interpretation of the colors is representing connected components in $F(X_i^{+})$ (the forest $F$ restricted to the vertices of $X_i^{+}$) and vertices in $X_i$ that have neighbors in $F$, i.e., vertices $u \in X_i$ with the same color $\kappa_i(u)=c \in [1, \omega-1]$ belong to the same connected component in the subsolution forest $F(X_i^{+})$. The vertices $u$ such that $\kappa_{i}(u)=\omega$ do not participate in the forest but have at least one neighbor in $F$. Finally, vertices $u$ with $\kappa_i(u)=0$ neither belong to the subsolution represented by $\kappa_i$ nor have a neighbor that belongs to it.
The idea behind this representation is the following. We assume that there is a Steiner tree $T$ in $G$. The forest $T(X_i^{+})$ is obtained by restricting the $1$-neighborhood of the tree $T$ to the vertices in $X_i^{+}$. This restriction of $N[T] \cap X_i^{+}$ is encoded by the configuration $\kappa_{i}$. In particular, for a Steiner tree $T$, there is a unique configuration $\kappa_i$ for each bag $X_i$ that is consistent with $T$ (as formally defined later). Overall, each table $X_i$ has at most $B_{k'}$ entries, for $k'=\omega+1$, where $B_k'=O(k'!)$ is the \emph{Bell-number} \cite{ChimaniMZ11}, corresponding to the number of partitions of at most $\omega$ elements into $\omega+1$ classes.
\vskip .1cm \noindent \textbf{Legal Configuration.}
Let $A(\kappa_i)=\{u \in X_i \mid \kappa_i(u)>0\}$ denote the set of active vertices in the configuration $\kappa_i$. A configuration $\kappa_i$ is \emph{legal} if it satisfies (L1) every terminal $u \in U_i$ is assigned a color in the range $\kappa_i(u) \in [1, \omega]$, and (L2) $N[A(\kappa_i)]\cap X_i \subseteq A(\kappa_i)$.
\par The algorithm proceeds by computing, for each legal configuration $\kappa_i \in tb(i)$, a value $Val(\kappa_i)$ that represents the minimum number of vertices in $X_i^{+}$ that belongs to the neighborhood of a forest $F$ that respects the configuration.
These values are computed bottom-up. Before we proceed, we first define the notion of compatible configurations.
\vskip .1cm \noindent \textbf{Compatible configurations.}
Let $i$ be parent of $j$ in the tree $\TD$.
Then $\kappa_i \in tb(i)$ is \emph{compatible} with $\kappa_j \in tb(j)$, denoted as  $\kappa_i \sim \kappa_j$ if the following four properties hold:\\
(Q1) $\kappa_i$ and $\kappa_j$ agree on their mutual vertices $X'=X_i \cap X_j$. For any $u \in X'$, $\kappa_i(u)=\omega$ iff $\kappa_j(u)=\omega$. In addition, for any pair $u,v \in X'$ that are in the same connected component, $\kappa_j(u)=\kappa_j(v) \in [1, \omega-1]$, it holds that $u$ and $v$ are connected in $\kappa_i$ as well. (Note that some components in the child $j$ might be connected in the parent $i$, and thus the converse does not hold.)
(Q2) If $\kappa_i$ connects two connected components of $\kappa_j$, then this is supported by the graph $G$ edges (i.e., the the edges required for the connectivity exist). \\
(Q3) Any terminal of $X_j$ is connected to some connected component in the forest represented by $\kappa_i$. Hence, $\kappa_i$ contains at most $|X_i|$ connected components and for each terminal $u \in X_j$ there exists some $v \in X_i \cap X_j$ such that $\kappa_j(v)=\kappa_j(u)$ and $\kappa_i(u)\in [1, \omega-1]$.\\
(Q4) $N[A(\kappa_j)] \cap X_i \subseteq A(\kappa_i)$.
\par The value $Val(\kappa_i)$ of the coloring $\kappa_i \in tb(i)$ is computed bottom-up. Let $X_i$ be a leaf. Then for every \emph{legal} $\kappa_i \in tb(i)$, set $Val(\kappa_i)=|A(\kappa_i)|$ and $Val(\kappa_i)=\infty$ if the configuration is invalid. Consider a non-leaf node $X_i$ with children $X_{i_1}$ and $X_{i_2}$. Let $\Psi=(\kappa_{i}, \kappa_{i_1}, \kappa_{i_2})$ be a triple of compatible configurations, such that $\kappa_{i_1} \in tb(i_1)$ and $\kappa_{i_2}\in tb(i_2)$.
Define
\begin{equation}
\label{eq:cost_tw_mid}
Val(\Psi)=Val(\kappa_{i_1}, \kappa_{i})+Val(\kappa_{i_2}, \kappa_{i})+|A(\kappa_{i})|~,
\end{equation}
where, for $j \in \{i_1,i_2\}$
\begin{equation}
\label{eq:cost_tw_mid_2}
Val(\kappa_{j}, \kappa_{i})=Val(\kappa_{j})-|A(\kappa_{i}) \cap A(\kappa_{j})|~.
\end{equation}
Since $(X_{i_1}^{+},X_{i_2}^{+},X_i)$ is a separation in the graph $G(X_i^{+})$, the correctness of Eq. (\ref{eq:cost_tw_mid}) follows from Eq. (\ref{eq:treewidth_cost}). Finally, for each entry $\kappa_i \in tb(i)$, maintain the minimum cost over any set of compatible triples.  Define,
\begin{equation}
\label{eq:alg_cost}
Val(\kappa_i)=\min_{\Psi=(\kappa_i, \kappa_{i_1},\kappa_{i_2})} \{Val(\Psi) \mid
\kappa_{i_1} \in tb(i_1),\kappa_{i_2} \in tb(i_2),
\kappa_{i_1} \sim \kappa_{i}, \kappa_{i_2} \sim \kappa_{i}\}~.
\end{equation}
%
%
%
The optimal solution value for the whole graph can be found in the root bag $X_{r}$ of $\TD$, identifying a cheapest solution where all vertices with color $\neq 0$ are contained in the same connected component (i.e., have the same color).
Computing the optimal solution (which is sufficiently represented by the set of nodes in the final tree) is possible by backtracking or by storing the set of edges for each row and each bag.
\subsubsection{Analysis}
For a given configuration entry $\kappa_i\in tb(i)$, the forest $F$ \emph{agrees} with $\kappa_i$ if the following four properties hold:\\
(Z1) $F(X_i^{+})$ has $C$ connected components where $C=|\{ \kappa(u) \in [1, \omega-1], u \in X_i\}|$  is the number of distinct colors in the coloring $\kappa_i$. \\
(Z2) All vertices $u \in X_i$ of the same color, $\kappa_i(u)$ are in the same connected component in $F(X_i^{+})$.\\
(Z3) The $X_i$ neighbors of the forest are active, $N[F]\cap X_i= A(\kappa_i)$.\\
(Z4) The terminals $U_i^{+}$ are connected in $F(X_i^{+})$ in the following manner.
Any terminal $u \in U_i^{+} \setminus X_i$ is in \emph{non-singleton} component in $F$. Specifically, it is connected to some node $v \in X_i$. In addition, the terminals of $U_i^{+} \cap X_i$, are in some connected component in $F$.
\par For a legal configuration $\kappa_i$ define the family of all forests in $G$ that agree with $\kappa_i$ as
\begin{equation}
\label{eq:forest_cost}
\mathcal{F}(\kappa_i)=\{ F \subseteq G \mid F \text{~agree with~} \kappa_i \}
\end{equation}
Let
\begin{equation}
\label{eq:val_forest}
Val(F, X_i)=|N[F] \cap X_i^{+}|
\end{equation}
and $Val^{*}(\kappa_i)=\min\{Val(F, X_i) \mid F \in \mathcal{F}(\kappa_i)\}$ be the minimum value of any forest $F \in \mathcal{F}(\kappa_i)$.
\begin{lemma}
\label{lem:cor}
$Val(\kappa_i)=Val^{*}(\kappa_i)$.
\end{lemma}
\Proof
This can be shown by a straightforward inductive proof on the decomposition tree. The base cases are leaf nodes where the hypothesis clearly holds. Consider $i\geq 1$ and assume the induction assumption holds for all descendants of bag $X_i$.
Let $F^{*}\in \mathcal{F}(\kappa_i)$ be the forest that attains the minimum value in the family $\mathcal{F}(\kappa_i)$, i.e., $Val(F^{*}, X_i) =Val^{*}(\kappa_i)$.
Let $X_{1}, X_{2}$ be the children of $X_i$ in $\TD$ and let $\kappa_{1} \in tb(1), \kappa_{2} \in tb(2)$ be the two compatible entries such that $F^{*} \in \mathcal{F}(\kappa_1) \cap \mathcal{F}(\kappa_2)$, hence $N[F^{*}] \cap X_j=A(\kappa_j)$ for $j \in \{1,2,i\}$.
For a forest $F \in \mathcal{F}(\kappa_j)$ and $j \in \{1,2,i\}$, define
\begin{equation}
\label{eq:delta_def}
\DIFF(F,X_j,X_i)=|N[F] \cap (X_j^{+} \setminus X_i)|.
\end{equation}

Since $(X_1^{+}, X_2^{+}, X_i)$ is a separation in $G(X_i^{+})$ and by Property (Q4) we have the following.

\begin{claim}
\label{cl:tw1}
$\DIFF(F,X_j,X_i)=Val(F, X_j)-|A(\kappa_i) \cap A(\kappa_j)|$ for every $F \in \mathcal{F}(\kappa_j)$, where $j \in \{1,2\}$.
\end{claim}

In addition, since $(X_1^{+}, X_2^{+}, X_i)$ is a separation in $G(X_i^{+})$ by Eq. (\ref{eq:treewidth_cost}), it follows that
\begin{equation}
\label{eq:tw_val_f_star}
Val(F^{*}, X_i)=\DIFF(F^{*},X_1,X_i)+\DIFF(F^{*},X_2,X_i)+|A(\kappa_i)|~.
\end{equation}
For $j \in \{1,2\}$, let $F_j^{*} \in \mathcal{F}(\kappa_j)$ be the forest that attains the minimum value of $Val(F_j^{*}, \kappa_j)$ in their family, i.e.,
$Val^{*}(\kappa_j)=Val(F_j^{*}, \kappa_j)$.
By induction assumption, $Val(\kappa_j)=Val(F_j^{*}, \kappa_j)$.
By Eq. (\ref{eq:cost_tw_mid_2}) and Claim \ref{cl:tw1}, we get that
\begin{eqnarray}
\label{eqn:tw_val12}
Val(\kappa_{j}, \kappa_{i})&=&Val(\kappa_{j})-|A(\kappa_{i}) \cap A(\kappa_{j})| \label{ineq:tw1}
\\&=&
Val(F_j^{*}, \kappa_j)-|A(\kappa_{i}) \cap A(\kappa_{j})| \label{ineq:tw2}
\\&=&
\DIFF(F_j^{*},X_j,X_i)~, \label{ineq:tw3}
\end{eqnarray}
where Ineq. (\ref{ineq:tw1}) follows from Eq. (\ref{eq:cost_tw_mid_2}),
Ineq. (\ref{ineq:tw2}) follows from induction assumption and
Ineq. (\ref{ineq:tw3}) follows from Claim \ref{cl:tw1}.
Thus since the triple $\Psi=(\kappa_i, \kappa_1, \kappa_2)$ consists of compatible configurations, by Eq.  (\ref{eq:cost_tw_mid}), Eq. (\ref{eq:cost_tw_mid_2}), Eq. (\ref{eq:alg_cost}) and Eq. (\ref{eqn:tw_val12})
we get that
\begin{equation}
\label{eq:alg}
Val(\kappa_i)\leq Val(\Psi)=\DIFF(F_1^{*},X_1,X_i)+\DIFF(F_2^{*},X_2,X_i)+|A(\kappa_i)|~.
\end{equation}
Assume for contradiction that $Val(\kappa_i) > Val^{*}(\kappa_i)$. Then, by
Eq. (\ref{eq:alg}) and Eq. (\ref{eq:tw_val_f_star}), we get that
there exists $j \in \{1,2\}$ such that
\begin{equation}
\label{ineq:delta}
\DIFF(F_j^{*},X_j,X_i)>\DIFF(F^{*},X_j,X_i).
\end{equation}
Without loss of generality let $j=1$.
Combining Claim \ref{cl:tw1} and Ineq. (\ref{ineq:delta}), we get that $Val(F^{*}, X_1)<Val(F^{*}_1, X_1)$, and since $F^{*} \in \mathcal{F}(\kappa_1)$, we end with contradiction to the induction assumption.
We therefore get that $Val(F^{*}, X_1)=Val(F^{*}_1, X_1)$ and $Val(F^{*}, X_2)=Val(F^{*}_2, X_2)$. The lemma follows.
\QED
We now complete the proof of Thm. \ref{thm:tw_correctness}
\Proof
Let $\Upsilon=\{\kappa_r \in tb(r) \mid \kappa_r(u) \in \{0,1,\omega\}, \forall u \in X_r\}$ be set of configurations in the root $r$ of $\TD$, in which all vertices are in a single connected component. Note that there exists a configuration $\kappa_r$
such that the optimal secluded tree $T^{*} \in \mathcal{F}(\kappa_r)$.  Let $\kappa_r^{*} \in \Upsilon$  be the configuration that attains the minimum value $Val(\kappa_r^{*})$ in the set. It then follows, by Lemma \ref{lem:cor}, $Val(\kappa_r^{*})=Val^{*}(\kappa_r^{*})=q^{*}(\Terminals)$.
The Steiner tree can be computed by a brute-force backtracking procedure.
\par We now turn to running time.
The size of any table $tb(i)$ is bounded by $O(\omega!\cdot \omega)$.
During the bottom-up traversal of $\TD$ we consider all possible row combinations for at most 3 tables. The merge operation can be done in $O(\omega)$ time, and hence requires overall $O(B_{\omega+1}^{3} \cdot \omega)$. All other operations for entry processing  are linear in the size of the bag, $O(\omega)$.
Note that computing the values of subsolution $\kappa_i$ are linear in the treewidth. Thus our time complexity is of the same as order as that of \cite{ChimaniMZ11}. In particular, it is linear for graphs with fixed treewidth, in this case the tree decomposition can also be computed in linear time and hence the total complexity is linear. If $TW(G)=O(\log n /\log \log n)$ then given the tree decomposition $\TD$ for $G$, the running time is $\widetilde{O}(n^{3})$.
\QED

Note that the algorithm extends also to directed and node-weighted graphs.
We now outline the modifications needed for these extensions.
The extension to weighted graph is immediate. The only modification needed
is to
define $\DIFF(A,S)$, $\DIFF(B,S)$ and $\Cost(T,S)$
according to the weighted cost as in Eq. (\ref{eq:wcost})
instead of the unweighted cost,
and apply a similar analysis.

The extension to directed graphs is a bit more involved.
Note that in the directed case it is not enough to store configurations that represent the different connected components since in order to construct a valid tree we need to make sure that every node has only one parent.
Therefore it is important to ``keep track'' of the roots of these connected components.
For each connected component $C$ we store as part of the configuration  the root $r(C)$ of this connected component.
Namely, we add to each configuration a root for each color.
In order to merge two connected components $C_1$ and $C_2$ to single tree with root $r(C_1)$, we only consider edges
from $r(C_2)$ to $C_1$. However, note that the root of a component does not necessarily have to be from the current bag and hence the number of different configurations might be too large. To overcome this issue we use the following observation.  Consider a bag $X_i$ and a configuration $\kappa_i \in tb(i)$.
Note that it cannot be that two different connected components in $\kappa_i$ would have both roots that are not from $X_i$. To see this, consider a subtree whose root $r$ is not in $X_i$, since $X_i$ is a separator then $r$ does not have edges to nodes in $V \setminus X_i^+$ forcing $r$ to be the root of the final tree. Therefore, there could be at most one such component. We thus restrict ourselves to configurations where there is at most one connected component with a root not in $X_i$ (otherwise the configuration is invalid). Notice that the number of valid configurations might increase by at most a factor of $n$ (for all options of the root not in $X_i$). Hence, the running time increases by a factor of $n$.
The rest of the analysis is similar and thus is omitted from this version.
\vskip .1cm \noindent \textbf{Bounded Density Graphs.}
Let $\DegCost^*(\Terminals)=\min \{\DegCost(T) \mid T \in \mathcal{T}(\Terminals)\}$.
We show the following.
\begin{proposition}
\label{prop:bounded_density}
Let $\cG$ be a hereditary class of graphs with a linear number of edges, i.e., a set of graphs such that for each $G \in \cG$, $|E(G)| \le \ell \cdot |V(G)|$ for some constant $\ell$, and where $G \in \cG$ implies that $G' \in \cG$ for each subgraph $G'$ of $G$.
Then $\DegCost^*(\Terminals) \leq 2\ell \cdot q^{*}(\Terminals)$.
\end{proposition}
\Proof
Note that for every Steiner tree $T$ in $G$, it holds that
\begin{equation}
\label{eq:bounded_den}
\Cost(T) \leq \DegCost(T) \leq 2\ell \cdot \Cost(T).
\end{equation}
The left inequality is immediate. We now prove the right inequality. Let $N=N[T]$ be the neighborhood of $T$ and let $G(N)$ be the subgraph of $G$ induced by $N$ and $E(N)$ be its edge set. We have that
\begin{eqnarray*}
\DegCost(T) =2|E(N)| \leq 2\ell |N| = 2\ell\cdot \Cost(T)~,
\end{eqnarray*}
where the first equality follows by definition, the second inequality follows from the hereditariness of $GG$ and the third equality follows by the definition of $\Cost(T)$.
Hence, Eq. \ref{eq:bounded_den} holds. In particular, for the optimal secluded tree $T^{*}$, we get that $$\Cost(T^{*})=q^{*}(\Terminals)\leq 2\ell \cdot \DegCost(T^{*}) \leq 2\ell \cdot \DegCost^*(\Terminals).$$ The proposition holds.
\QED
An example of such a class of graphs is the family of planar graphs. For this family, the above proposition yields a 6-approximation for the \PP\ problem. (A more careful direct analysis for planar graphs yields a 3-approximation, however) Finally, note that whereas computing $\DegCost^*(\Terminals)$ for a constant number of terminals $k$ is polynomial, for arbitrary $k$, computing $\DegCost^*(\Terminals)$ is NP-hard but can be approximated to within a ratio of $\Theta(\log k)$ thanks to Claim \ref{cl:degcost_approx}. Thus, for the class of bounded density graphs, by Prop. \ref{prop:bounded_density}, the \PP\ problem has a constant ratio approximation and the \PS\ problem has an $\Theta(\log k)$ ratio approximation.
\begin{theorem}
For the class of bounded-density graphs, the \PP\ (respectively, \PS) problem
is approximated within a ratio of $O(1)$ (resp., $\Theta(\log k)$).
\end{theorem}
%


\bigskip\bigskip\bigskip
{\small

}

\end{document}